\documentclass[11pt,a4paper,fleqn]{article}
\usepackage[a4paper,margin=2.2cm]{geometry}
\usepackage[utf8]{inputenc}
\usepackage[T1]{fontenc}
\usepackage{lmodern}
\usepackage[english]{babel}
\usepackage{appendix}
\usepackage{amsmath}
\usepackage{amsfonts}
\usepackage{amssymb}
\usepackage{graphicx}
\usepackage{booktabs}
\usepackage{adjustbox}
\usepackage{subcaption}
\usepackage{xspace}
\usepackage{afterpage}
\usepackage{setspace}

\newcommand{\eventfreegroup}{event-free group up to time $c$\xspace}
\newcommand{\eventgroup}{event group up to time $c$\xspace}

\usepackage[authoryear,round]{natbib}
\usepackage{titling}
\PassOptionsToPackage{linktocpage}{hyperref}
\usepackage[
  colorlinks=true,
  linkcolor=blue,
  citecolor=blue,
  urlcolor=blue,
]{hyperref}
\usepackage{breakcites}

\title{A Finite-Horizon Mixture Cure Model with Application to Online Flea Market Data}

\author{%
  \begin{tabular}{c}
    Yuji Komiyama\textsuperscript{1,*}
    \quad
    Yasumasa Matsuda\textsuperscript{2}
    \quad
    Masakazu Ishihara\textsuperscript{2,1}
    \\[0.4em]
    {\normalsize \textsuperscript{1}Graduate School of Economics and Management, Tohoku University}
    \\
    {\normalsize \textsuperscript{2}Leonard N. Stern School of Business, New York University}
    \\[0.4em]
    {\small \textsuperscript{*}Corresponding author: \texttt{komiyama.yuji.r2@dc.tohoku.ac.jp}}
  \end{tabular}%
}

\date{}

\begin{document}
\maketitle

\begin{abstract}
    This study proposes a mixture cure model that latently divides a population based on event occurrence within a finite time horizon. Conventional models rely on event occurrence over an infinite horizon, introducing untestable assumptions that often lead to issues with identifiability and interpretability. By shifting the estimand to a specific period of interest, the proposed approach reduces reliance on these infinite-tail assumptions and aligns interpretations more closely with finite-horizon decision-making objectives. Through simulation studies, we first evaluate the statistical properties of the proposed estimator, including estimation bias and variance. We further show that relying on conventional infinite-horizon models for finite-horizon decision-making can lead to erroneous judgments.  Finally, we apply the model to transaction data from Mercari, a Japanese online flea market platform. The empirical results reveal that the proposed model identifies different significant variables compared to the conventional model, offering interpretations that better reflect seasonal variation in user behavior.
\end{abstract}


\medskip

\section{Introduction}\label{sec:introduction}


In survival analysis, it is usually assumed that if a sufficiently long observation period is ensured for all study subjects, each individual will eventually experience the event of interest. Let $T$ denote the random variable representing the survival time and $F(t)$ its distribution function. The survival function, defined as the probability that the event has not occurred by time $t$, is given by
\begin{equation}
    S(t) = 1 - F(t).
\end{equation}
Standard survival models assume $\lim_{t \to \infty} S(t) = 0$. In practice, however, not all individuals necessarily experience the event. For example, in medical research where survival time is defined as the time from surgery to recurrence, some patients may be completely cured and never experience recurrence. Similarly, in economics, when survival time is defined as the time from job separation to reemployment, not everyone necessarily finds a new job. To address such situations in which not all individuals experience the event, cure models were introduced, allowing $\lim_{t \to \infty} S(t) > 0$. Individuals who never experience the event are regarded as \emph{cured}.

Cure models can be broadly classified into mixture cure models and promotion time cure models. A mixture cure model assumes that the population consists of two latent subgroups: \emph{a cured group} that will never experience the event, and \emph{a susceptible group} that will eventually experience it. A latent variable representing group membership is introduced. Let $S_{\infty}(t) = \mathbb{P}(T > t \mid T < \infty)$ denote the survival function of the susceptible group and $1-\pi_{\infty} = \mathbb{P}(T = \infty)$ the cure rate. The population survival function $S_{\text{pop}}(t)$ is then
\begin{equation}
    S_{\text{pop}}(t) = (1 - \pi_{\infty}) + \pi_{\infty} S_{\infty}(t),
\end{equation}
where $\lim_{t \to \infty} S_{\infty}(t) = 0$ is assumed. In contrast, the promotion time cure model proposed by \citet{1996Yakovlev_StochasticModelsTumor} is based on biological mechanisms and has a structure in which the same parameters determine both the cure rate and survival, making it impossible to separate these effects. For interpretability, the present study focuses on the mixture cure model. By allowing these quantities to depend on covariates $x$, the model separates two components of covariate effects: the \emph{incidence} component, which models whether an individual belongs to the susceptible group through $\pi_{\infty}(x) = \mathbb{P}(T < \infty \mid x)$, and the \emph{latency} component, which models the event-time distribution conditional on being susceptible through $S_{\infty}(t \mid x) = \mathbb{P}(T > t \mid x,\, T < \infty)$. Both components are defined in terms of whether the event ultimately occurs over an infinite time horizon.

In many applied settings, however, the question of whether the event \emph{ultimately} occurs is not the relevant question; decisions are instead governed by finite time horizons. A retailer deciding whether to discount or remove an unsold product needs to know whether it will sell within the next few months, not whether it would eventually sell given infinite time. A clinician evaluating a treatment is concerned with whether relapse occurs within a specific follow-up window. In each case, the finite-horizon answer is what drives the decision, and the infinite-horizon answer may be neither available nor necessary. For a given finite horizon $c > 0$, this question---whether the event occurs by time $c$, i.e., $\mathbb{P}(T < c \mid x)$---is a fundamentally different estimand from the infinite-horizon counterpart $\mathbb{P}(T < \infty \mid x)$, and the two need not agree. A covariate that reduces the ultimate event probability $\mathbb{P}(T < \infty \mid x)$ may simultaneously increase the short-term event probability $\mathbb{P}(T < c \mid x)$ for a given horizon $c$. Consequently, adopting an infinite-horizon model to inform finite-horizon decisions can lead to erroneous conclusions, yet this mismatch has received little attention in the literature.

A concrete instance of this mismatch arises in consumer-to-consumer (C2C) online marketplaces, where the seller is an individual consumer rather than a retailer and must physically store the listed item at home until the transaction is completed. The seller therefore bears holding costs---storage-space occupation, ongoing management effort, and the opportunity cost of alternative disposal channels such as secondhand shops, donation, or discarding---that accumulate with elapsed time, so the seller's welfare depends not on whether the item ever sells but on whether it sells within a period that the seller can tolerate holding it. The substantive estimand is then $\mathbb{P}(T < c \mid x)$ for a finite $c$ rather than $\mathbb{P}(T < \infty \mid x)$, making this setting a canonical case in which the finite-horizon framework developed in this study is required rather than merely convenient. We analyze this case in Section~\ref{sec:real-data-analysis} using transaction data from Mercari \citep{mercari2023}, one of the largest C2C online flea market platforms in Japan.

To address this gap, we shift the mixture cure framework itself---latent binary classification of the population with separate incidence and latency components---to a prespecified finite time horizon $[0, c)$, where $c$ is chosen by the analyst. The contributions of this study are threefold.
First, we develop a finite-horizon mixture cure framework in which the incidence component estimates $\mathbb{P}(T < c \mid x)$ and the latency component models the conditional event-time distribution on $[0, c)$. This framework enables joint estimation and interpretation of the probability of event occurrence by time $c$ and the conditional timing of the event within $[0,c)$, thereby providing a model suited to finite-horizon decision-making.
Second, by restricting the analysis to $[0, c)$, the model avoids reliance on assumptions about tail behavior beyond $c$: the event-time distribution on $(c,\infty)$ is left unspecified, and the possibility that the event occurs after $c$ is not ruled out. Incidence and latency interpretations therefore concern only behavior on $[0,c)$ and are explicitly contingent on the analyst's choice of $c$.
Third, through simulation studies we demonstrate that conventional infinite-horizon models can yield sign-reversed covariate effects relative to the finite-horizon truth, producing misleading guidance for finite-horizon decisions.

There exist numerous prior studies on mixture cure models. The mixture cure model was first proposed by \citet{1949Boag_MaximumLikelihoodEstimates,1952Berkson_SurvivalCurveCancer}, who considered models without covariates. Later, \citet{1977Farewell_ModelBinaryVariable} modeled the probability of not being cured using logistic regression. Furthermore, \citet{1992Kuk_MixtureModelCombining} extended the approach to allow the probability of not being cured to depend on covariates via logistic regression, and modeled the survival function for susceptible individuals with the Cox proportional hazards model \citep{1972Cox_RegressionModelsLifeTables}. Subsequent studies further developed estimation methods \citep{2000Peng_NonparametricMixtureModel,2000Sy_EstimationCoxProportional}. There are also extensions incorporating accelerated failure time (AFT) models \citep{2002Li_SemiparametricAcceleratedFailure} and nonparametric approaches \citep{2017Lopez-Cheda_NonparametricIncidenceEstimation}. This model is also extended to the economic field \citep{2019Dirick_MacroEconomicFactorsCredit} and the marketing field \citep{2018Kumar_AreYouBack}, among others.

To address the lack of information at infinite time, \citet{1995Taylor_SemiparametricEstimationFailure} introduced an assumption called the zero-tail constraint. Under this assumption, all censored observations with survival times longer than the maximum observed event time are treated as cured individuals. In other words, it is assumed that the follow-up period is sufficiently long. This assumption is reasonable when the survival function becomes flat at a non-zero value beyond a certain point, but in other cases it may lead to overestimation of the cure rate. To address this problem, \citet{2003Peng_EstimatingBaselineDistribution} proposed a model that, while expressing the survival function for susceptible individuals with the Cox proportional hazards model, introduces a decaying baseline survival function such as the Weibull or exponential distribution beyond the maximum observed event time. This approach aims to mitigate the influence of the zero-tail constraint. However, this method still relies on the assumption that survival times follow a certain decaying distribution, and does not fundamentally resolve identifiability issues. In recent years, various attempts have been made to relax this assumption. For example, to test whether truly cured individuals exist in the population when follow-up is limited, \citet{2019Escobar-Bach_NonParametricCureRate} proposed a new estimator for the cure rate using extrapolation techniques from extreme value theory. Tests for the sufficiency of follow-up duration \citep{2024PingXie_TestingSufficientFollowUp}, and nonparametric estimation methods based on extreme value theory \citep{2026Beirlant_NonparametricCureModels}, have also been developed. Additionally, \citet{2023Safari_LatencyFunctionEstimation} showed that when some cured individuals are known in advance (e.g., in settings where a medical test definitively verifies cure), including this information in estimation asymptotically reduces the variance of estimators. Compared to these previous works, the present study differs in that it shifts the mixture cure framework itself to the finite interval $[0, c)$. Rather than attempting to recover infinite-horizon quantities under limited follow-up, our model directly targets finite-horizon estimands---the probability of event occurrence by time $c$ and the conditional event-time distribution on $[0, c)$---ensuring identifiability through the boundary condition without reliance on assumptions about tail behavior beyond $c$. Accordingly, estimation of the cure rate is not the main objective in our model.

Literature that critically examines the practical interpretation of mixture cure models remains scarce. While \citet{2018Amico_CureModelsSurvival} notes that ``Indeed, the incidence models the long-term effect of covariates on the cure status, which is something permanent, whereas the latency focuses on the short-term, time-dependent effect that only concerns uncured observations,'' such discussions are largely confined to the infinite-horizon framework. If we analyze survival time for which longer duration is preferable (such as the time to recurrent cancer), the conventional interpretation may be useful. However, if we analyze the survival time for which shorter duration is preferable (such as the time to sale of a product), the conventional interpretation may not be useful. To the best of our knowledge, no existing studies have problematized the inconsistency between these conventional interpretations and the practical needs of finite-horizon decision-making. By addressing this gap, the proposed model offers a more practical interpretation than conventional models, particularly in scenarios where cured individuals exist but the primary analytical focus is on whether the event occurs within a finite time horizon.

The remainder of this paper is organized as follows. Section \ref{sec:methodology} introduces the proposed mixture cure model, providing mathematical definitions and justifications for the modeling framework. In Section \ref{sec:simulation-study}, we evaluate the desirable statistical properties of the proposed model, such as estimation bias and the standard deviation of estimates, using simulated data. Furthermore, we demonstrate that relying on conventional models—which divide the population based on infinite-horizon event occurrence—can lead to erroneous judgments in finite-horizon decision-making. Section \ref{sec:real-data-analysis} presents an application to transaction data from Mercari to analyze user behavior. The empirical analysis highlights that the proposed model offers more practical and intuitive interpretations, especially regarding the seasonality of products, and reveals differences in findings compared to conventional mixture cure models. Finally, Section \ref{sec:conclusion} summarizes the main findings and discusses their implications.

\section{Methodology}\label{sec:methodology}

In this section, we first define the mathematical notation required for the discussion of survival analysis. Then, we outline the proposed model, followed by a discussion on the estimation method.

\subsection{Preliminaries}\label{subsec:preliminaries}

For each individual $i=1,\dots,N$, let $T_i$ be the true survival time, $C_i$ be the right-censoring time, $t_i = \min(T_i, C_i)$ be the observed time, and $\delta_i = \mathbf{1}(T_i \le C_i)$ be the censoring indicator. Here, $\mathbf{1}(\cdot)$ is the indicator function. Let $T$ be the random variable representing the survival time, with its marginal distribution function $F(t)$ and survival function $S(t) = 1 - F(t)$. Let $x_{i} = (1, \tilde{x}_{i}) \in \mathcal{X} \subset \mathbb{R}^{p+1}$ be the covariate vector corresponding to each individual. Here, $1$ is the constant term, $\tilde{x}_{i} \in \mathbb{R}^{p}$ is the covariate vector excluding the constant term, and $\mathcal{X}$ represents the covariate vector space. Let $F(t \mid x)$ be the distribution function of the survival time conditional on $x$, and $S(t \mid x) = 1 - F(t \mid x)$ be the conditional survival function. Furthermore, we assume non-informative censoring, $T_{i} \perp C_{i} \mid x_{i}$. Thus, the observed data can be expressed as $\mathcal{D} = \{(x_{i}, \tilde{x}_{i}, t_i, \delta_i)\}_{i=1}^{N}$.

In the conventional mixture cure model \citep{2000Sy_EstimationCoxProportional}, it is assumed that the population is latently divided into two latent subgroups: a cured group in which the event will never occur in the future, and a susceptible group in which the event will occur at some point up to infinity. The population survival function $S_{\mathrm{pop}}$ is expressed as:
\begin{equation}
    S_{\mathrm{pop}}(t \mid x) = (1 - \pi_{\infty}(x)) + \pi_{\infty}(x) S_{\infty}(t \mid x), \quad t \in [0, \infty) \label{eq:mixture_cure_model}
\end{equation}
Here, $\pi_{\infty}(x) \in (0,1)$, and it is assumed that $\lim_{t \to \infty} S_{\infty}(t \mid x) = 0$. In this case, the cure rate is defined as $\lim_{t \to \infty} S_{\mathrm{pop}}(t \mid x) = 1 - \pi_{\infty}(x) = \mathbb{P}(T = \infty \mid X=x)$. The cure rate represents the probability that an individual with covariate $x$ will never experience the event in the future. $S_{\infty}(t \mid x) = \mathbb{P}(T > t \mid X=x, T < \infty)$ is the survival function of the susceptible group. In conventional models \citep{2000Sy_EstimationCoxProportional,2000Peng_NonparametricMixtureModel}, it is common to assume a logistic regression model for $\pi_{\infty}(x)$ and a Cox proportional hazards model for $S_{\infty}(t \mid x)$.

\subsection{Proposed Model}\label{subsec:proposal-model}

In this study, we classify the population based on the occurrence of the event at a finite time point $c \in (0, \infty)$, rather than whether the event occurs in infinite time. Here, $c$ is a prespecified value chosen by the analyst according to the decision-relevant time horizon, and the interpretation of the model differs depending on the chosen value of $c$. Specifically, the subpopulation in which the event does not occur by $t = c$ is regarded as \emph{the \eventfreegroup}, and the subpopulation in which the event occurs by $t = c$ is regarded as \emph{the \eventgroup}. The population survival function $S_{\mathrm{pop}}$ is expressed as:
\begin{equation}
    S_{\mathrm{pop}}(t \mid x) = (1 - \pi_{c}(x)) + \pi_{c}(x) S_{c}(t \mid x), \quad t \in [0, c) \label{eq:mixture_cure_model_time_c}
\end{equation}
Here, $\pi_{c}(x) \in (0,1)$. The survival function $S_{c}(t \mid x) = \mathbb{P}(T > t \mid X=x, T < c)$ of the \eventgroup is defined on $t \in [0, c)$, and we impose $\lim_{t \to c-} S_{c}(t \mid x) = 0$ so that $\lim_{t \to c-} S_{\mathrm{pop}}(t \mid x) = 1 - \pi_c(x) = \mathbb{P}(T \ge c \mid X=x)$ holds. $1 - \pi_{c}(x)$ represents the probability that an individual with covariate $x$ will not experience the event by $t=c$. It should be noted that the domain of the model changes between Equation \eqref{eq:mixture_cure_model} and Equation \eqref{eq:mixture_cure_model_time_c}. Since the proposed model restricts the domain to $t \in [0, c)$, no assumptions are made for $t \ge c$. Therefore, the possibility of the event occurring after $c$ is allowed, and individuals who actually experience the event after $c$ are classified into the \eventfreegroup. By restricting the analysis target to an observable and meaningful range in this way, clear estimation within the finite time horizon $[0, c)$ becomes possible without relying on the assumption of infinite time. As a result, practical interpretations, such as measuring the probability of event occurrence within period $c$ and the effect on the time to the event, become easier.

Let $f_{c}$ be the probability density function of the \eventgroup, and $f_{0c}$ be the baseline probability density function when $\tilde{x} = 0$.  Following \citet{2000Sy_EstimationCoxProportional}, we propose a model that assumes proportional hazards for the survival function of the \eventgroup. That is, we assume that the survival function $S_{c}(t \mid \tilde{x})$ of the \eventgroup is expressed in a Cox proportional hazards form:
\begin{equation}
    h_{c}(t \mid \tilde{x}; \beta) = h_{0c}(t)\exp(\tilde{x}^\top \beta)
    \label{eq:proportional_hazards_latency_model_hazard_function}
\end{equation}
Here, $h_{c}(t \mid \tilde{x}; \beta) = f_{c}(t \mid \tilde{x}; \beta) / S_{c}(t \mid \tilde{x}; \beta)$ is the hazard function, $h_{0c}(t) = f_{0c}(t) / S_{0c}(t)$ is the baseline hazard function, and $\exp(\tilde{x}^\top \beta)$ describes the effect of the covariate $\tilde{x}$. Note that to ensure the identifiability of the baseline survival function, the covariate vector $\tilde{x}$ does not include a constant term. Since $S_{c}(t \mid \tilde{x}; \beta) = \exp{\left(-\int_{0}^{t} h_{c}(u \mid \tilde{x}; \beta) \, du\right)}$, we can rewrite Equation \eqref{eq:proportional_hazards_latency_model_hazard_function} as:
\begin{equation*}
    S_{c}(t \mid \tilde{x}; \beta) = S_{0c}(t)^{\exp(\tilde{x}^\top \beta)}
\end{equation*}
where $S_{0c}(t) = \exp{\left(-\int_{0}^{t} h_{0c}(u) \, du\right)}$ is the baseline survival function.

Because the conditional event-time distribution of the \eventgroup is supported on the finite interval $[0, c)$, the boundary condition $\lim_{t \to c-} S_{0c}(t) = 0$ must hold, which in turn forces $\lim_{t \to c-} h_{0c}(t) = \infty$. Directly modeling such a divergent baseline hazard is difficult. We therefore take the baseline density $f_{0c}$ as the primary object of nonparametric modeling on $[0, c)$, and induce the baseline survival function and the baseline hazard from it via
\begin{equation*}
    S_{0c}(t) = 1 - \int_{0}^{t} f_{0c}(u) \, du, \qquad h_{0c}(t) = \frac{f_{0c}(t)}{S_{0c}(t)}.
\end{equation*}
Since $f_{0c}$ is a probability density on $[0, c)$ by construction, the boundary condition $\lim_{t \to c-} S_{0c}(t) = 0$ holds by definition.

A key advantage of this density-based construction is that the standard Cox interpretation of the regression coefficient $\beta$ is preserved on $[0, c)$. Although the induced $h_{0c}(t)$, and hence $h_{c}(t \mid \tilde{x}; \beta)$, diverges as $t \to c-$, the hazard ratio between any two covariate values $\tilde{x}_{1}$ and $\tilde{x}_{2}$ is
\begin{equation*}
    \frac{h_{c}(t \mid \tilde{x}_{1}; \beta)}{h_{c}(t \mid \tilde{x}_{2}; \beta)} = \exp{\left((\tilde{x}_{1} - \tilde{x}_{2})^{\top} \beta\right)},
\end{equation*}
which is finite and time-independent because the divergent baseline cancels in the ratio. Hence $\beta$ retains exactly the same interpretation as in the standard Cox proportional hazards regression \citep{1972Cox_RegressionModelsLifeTables}, and the proposed model reconciles the boundary divergence required by the finite-horizon constraint with the familiar covariate-effect interpretation of the Cox model.

To realize this design, we represent the baseline density $f_{0c}$ as a mixture of normalized cubic B-spline basis functions $\{\tilde{B}_{i,3}(t)\}_{i=1}^{K}$. Following \citet{deBoor1978}, let $B_{i,3}(t)$ denote the standard cubic B-spline basis function defined by a knot sequence. We define the normalized basis as
\begin{equation*}
    \tilde{B}_{i,3}(t) = \frac{B_{i,3}(t)}{\int_{0}^{c} B_{i,3}(u) \, du}
\end{equation*}
so that $\int_{0}^{c} \tilde{B}_{i,3}(t) \, dt = 1$ holds. Cubic B-splines have the property of being able to uniformly approximate smooth curves if the knots are appropriately set. Moreover, since $B_{i,3}(t)$ is a piecewise polynomial, the integral $\int_{0}^{t} \tilde{B}_{i,3}(u) \, du$ in \eqref{eq:proportional_hazards_latency_model_survival_function} can be written in closed form, which facilitates efficient computation.

Concretely, we define the baseline density on $[0, c)$ as the linear combination
\begin{align*}
    f_{0c}(t; \alpha) & = \sum_{i=1}^{K} \gamma_i(\alpha) \tilde{B}_{i,3}(t), \\ \gamma_i(\alpha) & = \frac{\exp(\alpha_i)}{\sum_{j=1}^{K} \exp(\alpha_j)}, \quad \alpha_K = 0
\end{align*}
Here, $\alpha = (\alpha_{1}, \ldots, \alpha_{K})$ (where $\alpha_K=0$ is fixed to avoid unidentifiability arising from the shift invariance of the softmax function), $\gamma_i(\alpha) > 0$ for all $i$ (since the softmax function is strictly positive), and $\sum_{i=1}^{K} \gamma_i(\alpha) = 1$. Thus $f_{0c}(\cdot; \alpha)$ is a valid probability density on $[0, c)$ with $\lim_{t \to c-} \int_{0}^{t} f_{0c}(u; \alpha) \, du = 1$. The baseline survival function induced from $f_{0c}$ is
\begin{equation}
    S_{0c}(t; \alpha) = 1 - \sum_{i=1}^{K} \gamma_i(\alpha) \int_{0}^{t} \tilde{B}_{i,3}(u) \, du
    \label{eq:proportional_hazards_latency_model_survival_function}
\end{equation}
for $t \in [0, c)$. Then $S_{0c}(t; \alpha) > 0$ for all $t \in [0, c)$ and $\lim_{t \to c-} S_{0c}(t; \alpha) = 0$. This ensures that the Cox-type density
\begin{align*}
    f_{c}(t \mid \tilde{x}; \beta, \alpha) & = -\frac{d}{dt} S_{c}(t \mid \tilde{x}; \beta, \alpha) \\ & = \exp(\tilde{x}^\top \beta) f_{0c}(t; \alpha) S_{0c}(t; \alpha)^{\exp(\tilde{x}^\top \beta) - 1}
\end{align*}
is well-defined on $[0, c)$; in particular, when $\exp(\tilde{x}^\top \beta) - 1 < 0$, the factor $S_{0c}(t; \alpha)^{\exp(\tilde{x}^\top \beta) - 1}$ remains finite because $S_{0c}(t; \alpha) > 0$ on the defined interval.
The knot sequence fixes the endpoints of $[0,c]$, extends with equally spaced knots outside the interval, and places internal knots at the empirical quantiles of the observed event times. The resulting latency model can be viewed as a sieve-based proportional hazards model \citep{1972Cox_RegressionModelsLifeTables} on the finite interval $[0, c)$: the Cox proportional hazards structure governs covariate effects, while the baseline density is approximated nonparametrically as a B-spline mixture, from which the baseline survival and hazard functions are induced.

We assume a logistic regression model for $\pi_c(x)$:
\begin{equation}
    \pi_c(x; b) = \frac{1}{1 + \exp(-x^\top b)}
    \label{eq:event_probability}
\end{equation}
Here, $b = (b_{1}, \ldots, b_{p+1})$ is the parameter vector of the logistic regression model. Therefore, the parameters to be estimated are $(b, \beta, \alpha)$, and the total number is $2p + K(= (p+1) + p + (K-1))$.

\subsection{Estimation Method}\label{subsec:estimation-method}

Following \citet{2000Sy_EstimationCoxProportional}, we introduce a latent variable $z_{i}$ representing membership in the \eventfreegroup or the \eventgroup, and perform parameter estimation using the EM algorithm. Here, if $z_{i}=0$, the $i$-th individual belongs to the \eventfreegroup, and if $z_{i}=1$, they belong to the \eventgroup. Therefore, each data point is represented by $\{(\tilde{x}_i, x_{i}, t_i, \delta_i, z_{i})\}_{i=1}^{N}$.
When $t_i < c$, if $\delta_i = 1$, the individual belongs to the \eventgroup, and if $\delta_i = 0$, they belong to either the \eventfreegroup or the \eventgroup. On the other hand, when $t_i \ge c$, all are considered to belong to the \eventfreegroup. Based on the above, the observed log-likelihood based on the observed data is expressed as:
\begin{align*}
    \ell(b, \beta, \alpha)
     & = \sum_{i: t_i < c} \delta_i \log\big(\pi_{c}(x_i; b) f_{c}(t_i \mid \tilde{x}_i; \beta, \alpha)\big)                           \\
     & + \sum_{i: t_i < c} (1-\delta_i) \log\big((1-\pi_{c}(x_i; b)) + \pi_{c}(x_i; b) S_{c}(t_i \mid \tilde{x}_i; \beta, \alpha)\big) \\
     & + \sum_{i: t_i \ge c} \log(1-\pi_{c}(x_i; b))
\end{align*}
In the softmax function used in the proposed model, there is near-unidentifiability with respect to $\{\alpha_{i}\}_{i=1}^{K-1}$. In other words, when a certain $\alpha_{i}$ takes a large value, slightly changing $\alpha_{j}$ ($j \neq i$) may hardly change the value of $\ell(b, \beta, \alpha)$. As a result, the negative Hessian matrix of $\ell(b, \beta, \alpha)$ may not be a positive definite matrix. Therefore, estimation by Newton's method can become unstable. To prevent this, we specify a prior distribution for $\{\alpha_{i}\}_{i=1}^{K-1}$. Namely, we assume a normal distribution $\alpha \sim \mathcal{N}(0, \lambda^{-1} I_{K-1})$. Here, $\lambda$ is a regularization parameter (hyperparameter). This corresponds to L2 regularization. We consider MAP estimation.
\begin{align*}
    \hat{\theta}_{\mathrm{MAP}} & = \operatorname*{argmax}_{\theta} p(\theta \mid \mathcal{D})                                                \\
                                & = \operatorname*{argmax}_{\theta} \left( \log{p(\mathcal{D} \mid \theta)} + \log{p(\theta|\lambda)} \right) \\
                                & = \operatorname*{argmax}_{\theta} \left( \ell(b, \beta, \alpha) - \frac{\lambda}{2} ||\alpha||^{2} \right)
\end{align*}
The complete-data log-likelihood is
\begin{align}
    \ell_{C}(b, \beta, \alpha; z)
     & = \ell_{1}(b; z) + \ell_{2}(\beta, \alpha; z),
    \label{eq:complete_log_likelihood}
    \\
    \ell_{1}(b; z)
     & = \sum_{i=1}^{N} \left\{ z_{i} \log{(\pi_{c}(x_{i}; b))} + (1-z_{i}) \log{(1 - \pi_{c}(x_{i}; b))} \right\},                                                                                                        \nonumber \\
    \ell_{2}(\beta, \alpha; z)
     & = \sum_{i=1}^{N} \left\{ z_{i}\delta_{i} \log{f_{c}(t_{i}\mid \tilde{x}_{i}; \beta, \alpha)} + z_{i}(1- \delta_{i}) \log{S_{c}(t_{i}\mid \tilde{x}_{i}; \beta, \alpha)} \right\} - \frac{\lambda}{2} \|\alpha\|^{2}
    \nonumber
\end{align}
Let $\theta^{(m)} = (b^{(m)}, \beta^{(m)}, \alpha^{(m)})$ be the parameters at the $m$-th iteration. In the E-step, $w_i^{(m)}$ is calculated as follows:
\begin{align*}
    w_{i}^{(m)} = \mathbb{E}[Z_{i}|\theta^{(m)}, \mathcal{D}] =  \begin{cases}
                                                                     1,                                                                                                                                                                                                         & \text{if $\delta_{i} = 1$ and $t_{i} < c$,} \\
                                                                     \left. \dfrac{\pi_{c}(x_{i}; b) S_{c}(t_{i} \mid \tilde{x}_{i}; \beta, \alpha)}{1 - \pi_{c}(x_{i}; b) + \pi_{c}(x_{i}; b) S_{c}(t_{i} \mid \tilde{x}_{i}; \beta, \alpha)} \right|_{\theta = \theta^{(m)}}, & \text{if $\delta_{i} = 0$ and $t_{i} < c$,} \\
                                                                     0,                                                                                                                                                                                                         & \text{if $t_{i} \ge c$.}                    \\
                                                                 \end{cases}
\end{align*}
In the conventional model \citep{2000Sy_EstimationCoxProportional}, the E-step conditions only on $\delta_i$: if $\delta_i = 1$, the individual is known to be susceptible; if $\delta_i = 0$, the membership remains ambiguous. In contrast, the proposed model introduces the condition $t_i \ge c$, under which $w_i^{(m)} = 0$ is assigned deterministically. This yields a key advantage: individuals with follow-up times beyond $c$ are definitively classified as not having experienced the event by time $c$, thereby extracting more information from the data than the conventional formulation.
By substituting this into Equation \eqref{eq:complete_log_likelihood}, we obtain the expected log-likelihood $\tilde{\ell}_{C}(b, \beta, \alpha; w^{(m)}) = \tilde{\ell}_{1}(b; w^{(m)}) + \tilde{\ell}_{2}(\beta, \alpha; w^{(m)})$, where $w^{(m)} = (w_{i}^{(m)}; i=1,\ldots,N)^\top$. In the M-step, we maximize $\tilde{\ell}_{C}$ with respect to $b, \beta$, and $\alpha$. Since $\tilde{\ell}_{C}$ is expressed as the sum of $\tilde{\ell}_{1}$ and $\tilde{\ell}_{2}$, and the parameters are partitioned between the two terms, maximizing each of them will maximize $\tilde{\ell}_{C}$. They are each estimated using a quasi-Newton method. By the Laplace approximation, the posterior distribution $p(\theta|\mathcal{D})$ can be approximated by a Gaussian distribution with mean $\hat{\theta}_{\mathrm{MAP}}$ and precision matrix $A$ around $\theta = \hat{\theta}_{\mathrm{MAP}}$, where
\begin{equation*}
    A = -\nabla^2_\theta \left( \ell(b, \beta, \alpha) - \frac{\lambda}{2} \|\alpha\|^2 \right)\Big|_{\theta = \hat{\theta}_{\mathrm{MAP}}}.
\end{equation*}
Using this, we calculate the Bayesian credible intervals.
The hyperparameter $\lambda$ is estimated by the empirical Bayes method \citep{MacKay1992BayesianInterpolation}.
The estimation procedure is as follows (see Appendix~\ref{sec:empirical-bayes-method} for details).
Let $m = 1,2,\ldots$ index the outer iterations of empirical Bayes updates, let $\lambda^{(0)} > 0$ be an initial value for $\lambda$, and set $\hat{\lambda}^{(0)} = \lambda^{(0)}$.
\begin{enumerate}
    \item Given $\hat{\lambda}^{(m-1)}$, perform MAP estimation with hyperparameter $\hat{\lambda}^{(m-1)}$ to obtain $\hat{\theta}_{\mathrm{MAP}}$.
    \item Using $\hat{\theta}_{\mathrm{MAP}}$, find all roots $\lambda^{(m)}_{j}$ ($j = 1,\ldots,J_{m}$) of $g(\lambda) = 0$ such that
          \begin{align*}
              g(\lambda) = \sum_{i=1}^{M_{\alpha}} \frac{\mu_{i}}{\mu_{i} + \lambda} - \lambda \|\hat{\alpha}\|^2 = 0, \quad g'(\lambda^{(m)}_{j}) < 0
          \end{align*}
          where $M_{\alpha} = K - 1$ is the dimension of $\alpha$, $H_{xy} = -\nabla^2_{xy}\log p(\mathcal{D}|\theta)|_{\theta = \hat{\theta}}$ for $x,y \in \{\tilde\theta, \alpha\}$ with $\tilde{\theta} = (b, \beta)$, and $\mu_i$ ($i=1,\ldots,M_{\alpha}$) are the eigenvalues of $S := H_{\alpha\alpha} - H_{\alpha\tilde\theta} H_{\tilde\theta\tilde\theta}^{-1} H_{\tilde\theta\alpha}$.
          Set $\hat{\lambda}^{(m)} = \operatorname*{argmax}_{\lambda \in \{\lambda^{(m)}_{1},\ldots,\lambda^{(m)}_{J_{m}}\}} \log p(\mathcal{D}|\lambda)$.
    \item Repeat the above steps. If
          \begin{equation*}
              \left| \log p(\mathcal{D}|\hat{\lambda}^{(m)}) - \log p(\mathcal{D}|\hat{\lambda}^{(m-1)}) \right| < \epsilon
          \end{equation*}
          for a certain threshold $\epsilon > 0$, the procedure is regarded as converged.
    \item Using the converged $\hat{\lambda}^{(m)}$, perform MAP estimation again and take the resulting $\hat{\theta}_{\mathrm{MAP}}$ as the final estimator.
\end{enumerate}

\section{Simulation Studies}\label{sec:simulation-study}

In this section, we evaluate the proposed model using simulation studies and demonstrate the potential pitfalls of conventional models in finite-horizon settings.

\subsection{Scenario A}\label{subsec:scenario-a}
In Scenario A, we evaluate the performance of the proposed model using synthetic data. The proposed model serves as the true data-generating mechanism. Although the model is estimated via Bayesian MAP inference, we evaluate its performance using frequentist criteria. We generate synthetic data given sample size $N$, number of covariates $p$, and $c$, and apply the proposed EM algorithm and variance estimation based on the Laplace approximation. We evaluate the performance using empirical bias, coverage probability (CP), and credible interval width for the regression coefficients.
The covariate vector $\tilde{x}_i$ combines $p_\mathrm{cont}$ continuous covariates drawn from $N(0,1)$ and categorical covariates (one-hot encoded). We set $p_\mathrm{cont} = 1$ and three categorical variables with 4, 3, and 2 levels, yielding $p = 7$, to mirror the predominantly categorical structure of the real data in Section~\ref{sec:real-data-analysis}.
The vector $x_i = (1, \tilde{x}_i^\top)^\top$ is fixed across all replications. In each replication, the latent variable $z_i$ is generated as $z_i \sim \mathrm{Bernoulli}\bigl(\pi_{c}(x_i; b)\bigr)$ based on the logistic regression model \eqref{eq:event_probability}. The coefficient vector is set to $b = (b_1, \dots, b_{p+1})^\top$, where the intercept $b_1$ controls the overall proportion of the \eventgroup. The true value of the coefficients except for $b_{1}$ is $(b_{2}, \dots, b_{p+1}) = (-0.3,\ 0.5,\ 0.4,\ 0.2,\ 0.0,\ -0.2,\ -0.5)^\top$.

For individuals with $z_i = 1$, the event time $T_i$ is generated from the survival function $S_c(t \mid \tilde{x}_i; \beta)$ using the inverse transform sampling. For individuals with $z_i = 0$, the event time is set to $T_i = c + U_i$, where $U_i \sim \mathrm{Exponential}(\lambda_{>c})$ with $\lambda_{>c} = 0.05$. This setting allows us to evaluate model performance when events for the "cured" group technically occur after $c$, but are unobserved within the finite window.
For the baseline survival distribution of the \eventgroup, we adopt a distribution on the finite interval $[0, c)$ with shape parameter $\eta = 1.5$:
\begin{align*}
    S_{0c}(t) = 1 - (t/c)^{\eta}, \quad 0 \le t < c.
\end{align*}
Using this baseline in the Cox proportional hazards form yields the conditional survival function for data generation. The estimated model approximates this baseline via the B-spline representation in Equation~\eqref{eq:proportional_hazards_latency_model_survival_function}.
The true value of the coefficients for the latency (survival) model is $\beta = (0.3,\ -0.4,\ -0.2,\ 0.0,\ 0.2,\ 0.4,\ 0.5)^\top$.

Independent censoring times $C_i$ are generated from an exponential distribution with rate $\lambda_{\mathrm{cens}}$. We consider two sub-scenarios (A-1, A-2) by varying the \eventgroup proportion and censoring rate.
\begin{enumerate}
    \item Scenario A-1 represents a high \eventgroup proportion (70\%) with a standard censoring rate (30\% among $z_i=1$).
    \item Scenario A-2 represents a low \eventgroup proportion (30\%) with a standard censoring rate (30\% among $z_i=1$).
\end{enumerate}
The parameters $b_1$ and $\lambda_{\mathrm{cens}}$ were tuned using a monotone line search on a pilot dataset of $N=100,000$ to achieve these target proportions. The final values are: Scenario A-1 ($b_{1} = 0.928, \lambda_{\mathrm{cens}} = 0.06$); Scenario A-2 ($b_{1} = -0.838, \lambda_{\mathrm{cens}} = 0.06$).

We perform simulations with sample sizes $N = 500, 1000$, setting the number of replications $M=500$, $K=7$, and $c=10$. We evaluate the estimation accuracy of $\beta$ and $b$ using bias, coverage probability, and credible interval width. Additionally, the baseline survival function is evaluated using the Root Mean Integrated Squared Error (RMISE):
\begin{align*}
    \mathrm{RMISE} & = \sqrt{ \frac{1}{M} \sum_{m=1}^M \left[ \frac{1}{c} \int_0^c \left( \hat{S}_{0c, m}(t) - S_{0c}(t) \right)^2 dt \right] }. \\
\end{align*}
Here, $\hat{S}_{0c, m}(t)$ denotes the estimated baseline survival function in the $m$-th replication, and $S_{0c}(t)$ is the true baseline survival function. The integral is evaluated numerically using the trapezoidal rule. Specifically, the interval $[0, c]$ is divided into $J$ equally spaced grid points $t_1, \ldots, t_J$ with $t_{j+1} - t_j = c / J$; in this simulation, we set $J = 1000$.
The simulation results are summarized in Tables \ref{tab:simulation_for_proposed_model_summary} and \ref{tab:rmise_summary_scenario_sample}. As shown in Table \ref{tab:simulation_for_proposed_model_summary}, both scenarios A-1 and A-2 demonstrate a tendency for the bias and standard deviation to decrease as the sample size increases. The coverage probabilities are mostly close to 95\%, which are near the nominal level. Furthermore, transitioning from scenario A-1 to A-2, the bias and standard deviation of the parameters in the incidence part decrease slightly, while those in the latency part increase. This behavior is consistent with the following mechanism: a decrease in the event rate up to time $c$ leads to an increased proportion of the \eventfreegroup. Consequently, the number of individuals with an observation period greater than $c$ increases, which in turn increases the number of individuals with $w^{(m)}_{i}=0$, thereby slightly stabilizing the estimation for the incidence part. At the same time, this decrease in the event rate reduces the effective sample size available for estimating the latency part. Additionally, Table~\ref{tab:rmise_summary_scenario_sample} shows that for both A-1 and A-2, the RMISE decreases as the sample size increases, whereas transitioning from A-1 to A-2, the RMISE increases as the event rate decreases. The former observation is consistent with an increase in the effective sample size for estimating the baseline function as the overall sample size grows. The latter is consistent with a reduction in the effective sample size for baseline function estimation due to the decreased event rate. Based on these findings, it is suggested that the proposed model performs the estimations correctly. Additionally, we conducted a sensitivity analysis by varying the number of basis functions $K \in \{5, 7, 10, 15\}$. The estimated regression coefficients remained largely unchanged across these choices, indicating that the proposed model is robust with respect to the specification of $K$.

\begin{table}[p]
    \centering
    \caption{Simulation results for proposed model. Comparison of empirical bias (Bias), empirical standard deviation (SD), 95\% coverage probability (CP), and mean 95\% credible interval width (Width) across all scenarios ($N=500$ vs $N=1000$, $M=500$ replications).}
    \label{tab:simulation_for_proposed_model_summary}

    \begin{adjustbox}{max width=\textwidth, max totalheight=0.95\textheight, keepaspectratio}
        \small
        \renewcommand{\arraystretch}{0.9}

        \begin{tabular*}{\textwidth}{@{\extracolsep{\fill}} l c rrrr rrrr }
            \toprule
            & & \multicolumn{4}{c}{\textbf{$N=500$}} & \multicolumn{4}{c}{\textbf{$N=1000$}} \\
            \cmidrule(lr){3-6} \cmidrule(lr){7-10}
            \textbf{Param} & \textbf{True} & \textbf{Bias} & \textbf{SD} & \textbf{CP} & \textbf{Width} & \textbf{Bias} & \textbf{SD} & \textbf{CP} & \textbf{Width} \\
            \midrule

            \multicolumn{10}{l}{\textbf{Scenario A-1: High Event Rate (70\%)}}                                                                                                                                                              \\
            \multicolumn{10}{l}{\textit{Incidence Model Parameters (Logistic)}}                                                                                                                                                             \\
            $b_1$      &  0.928 &  0.024 & 0.343 & 0.952 & 1.315 &  0.027 & 0.236 & 0.960 & 0.932 \\
            $b_2$ & -0.300 & -0.014 & 0.134 & 0.966 & 0.540 & -0.010 & 0.092 & 0.950 & 0.366 \\
            $b_3$&  0.500 & -0.000 & 0.380 & 0.942 & 1.451 &  0.011 & 0.265 & 0.952 & 0.999 \\
            $b_4$&  0.400 &  0.005 & 0.366 & 0.952 & 1.458 &  0.003 & 0.266 & 0.938 & 1.012 \\
            $b_5$&  0.200 &  0.012 & 0.360 & 0.942 & 1.381 &  0.002 & 0.243 & 0.944 & 0.948 \\
            $b_6$&  0.000 &  0.001 & 0.328 & 0.940 & 1.253 & -0.018 & 0.234 & 0.936 & 0.881 \\
            $b_7$& -0.200 & -0.005 & 0.333 & 0.952 & 1.249 & -0.035 & 0.224 & 0.940 & 0.865 \\
            $b_8$& -0.500 & -0.020 & 0.262 & 0.958 & 1.027 & -0.010 & 0.192 & 0.944 & 0.714 \\

            \addlinespace[2pt]
            \multicolumn{10}{l}{\textit{Latency Model Parameters (Survival)}}                                                                                                                                                               \\
            $\beta_1$ &  0.300 &  0.002 & 0.078 & 0.932 & 0.290 &  0.003 & 0.050 & 0.946 & 0.196 \\
            $\beta_2$& -0.400 &  0.034 & 0.179 & 0.930 & 0.678 &  0.019 & 0.121 & 0.954 & 0.489 \\
            $\beta_3$& -0.200 &  0.037 & 0.181 & 0.946 & 0.722 &  0.024 & 0.132 & 0.940 & 0.512 \\
            $\beta_4$&  0.000 &  0.037 & 0.182 & 0.954 & 0.735 &  0.028 & 0.125 & 0.940 & 0.501 \\
            $\beta_5$&  0.200 &  0.026 & 0.157 & 0.958 & 0.626 &  0.024 & 0.105 & 0.968 & 0.445 \\
            $\beta_6$&  0.400 &  0.029 & 0.165 & 0.952 & 0.664 &  0.020 & 0.112 & 0.958 & 0.461 \\
            $\beta_7$&  0.500 &  0.019 & 0.140 & 0.952 & 0.553 &  0.010 & 0.099 & 0.940 & 0.384 \\
            \midrule

            \multicolumn{10}{l}{\textbf{Scenario A-2: Low Event Rate (30\%)}} \\
            \multicolumn{10}{l}{\textit{Incidence Model Parameters (Logistic)}} \\
            $b_1$          & -0.838 & -0.020 & 0.311 & 0.968 & 1.286 & -0.001 & 0.230 & 0.960 & 0.907 \\
            $b_2$     & -0.300 & -0.013 & 0.138 & 0.938 & 0.516 & -0.008 & 0.090 & 0.942 & 0.352 \\
            $b_3$    &  0.500 &  0.015 & 0.357 & 0.954 & 1.377 & -0.013 & 0.253 & 0.950 & 0.961 \\
            $b_4$    &  0.400 &  0.010 & 0.350 & 0.958 & 1.418 & -0.007 & 0.270 & 0.938 & 0.988 \\
            $b_5$    &  0.200 & -0.017 & 0.361 & 0.962 & 1.425 & -0.010 & 0.243 & 0.948 & 0.972 \\
            $b_6$    &  0.000 & -0.014 & 0.300 & 0.942 & 1.169 & -0.002 & 0.211 & 0.962 & 0.829 \\
            $b_7$    & -0.200 & -0.005 & 0.310 & 0.952 & 1.204 & -0.003 & 0.203 & 0.958 & 0.838 \\
            $b_8$    & -0.500 & -0.001 & 0.250 & 0.948 & 0.983 &  0.000 & 0.178 & 0.948 & 0.683 \\
            \addlinespace[2pt]
            \multicolumn{10}{l}{\textit{Latency Model Parameters (Survival)}} \\
            $\beta_1$ &  0.300 &  0.003 & 0.112 & 0.962 & 0.462 &  0.003 & 0.077 & 0.942 & 0.308 \\
            $\beta_2$& -0.400 &  0.070 & 0.262 & 0.966 & 1.035 &  0.062 & 0.186 & 0.952 & 0.748 \\
            $\beta_3$& -0.200 &  0.094 & 0.294 & 0.936 & 1.123 &  0.056 & 0.203 & 0.942 & 0.790 \\
            $\beta_4$&  0.000 &  0.089 & 0.299 & 0.942 & 1.190 &  0.069 & 0.201 & 0.956 & 0.788 \\
            $\beta_5$&  0.200 &  0.066 & 0.229 & 0.962 & 0.976 &  0.019 & 0.176 & 0.942 & 0.686 \\
            $\beta_6$&  0.400 &  0.050 & 0.271 & 0.954 & 1.061 &  0.035 & 0.185 & 0.944 & 0.722 \\
            $\beta_7$&  0.500 &  0.029 & 0.243 & 0.936 & 0.891 &  0.022 & 0.147 & 0.962 & 0.607 \\

            \bottomrule
        \end{tabular*}
    \end{adjustbox}
\end{table}

\begin{table}[p]
    \centering
    \caption{RMISE summary by scenario and sample size.}
    \begin{tabular}{l r}
        \toprule
        \textbf{Scenario}        & \textbf{RMISE} \\ \midrule
        Scenario A-1 ($N=500$):  & 0.040987       \\
        Scenario A-1 ($N=1000$): & 0.029934       \\
        Scenario A-2 ($N=500$):  & 0.058478       \\
        Scenario A-2 ($N=1000$): & 0.044473       \\
        \bottomrule
    \end{tabular}
    \label{tab:rmise_summary_scenario_sample}
\end{table}

\subsection{Scenario B}\label{subsec:scenario-b}
In this subsection, we show that when decisions are made over a finite horizon, a model that assumes a different structure---such as the conventional infinite-horizon cure model---can yield misleading conclusions and lead to incorrect decisions. Using simulation data, we verify that the proposed finite-horizon model correctly answers the two questions of interest: ``What is the probability that the event occurs within the $c$-period?'' and ``What is the time until the event occurs among those who experience it within the $c$-period?''

\begin{figure}
    \centering
    \subcaptionbox{Kaplan–Meier Estimator (Susceptible Only)\label{fig:scenario-b-only-susceptible-km}}{%
        \includegraphics[width=0.48\linewidth, height=0.22\textheight, keepaspectratio]{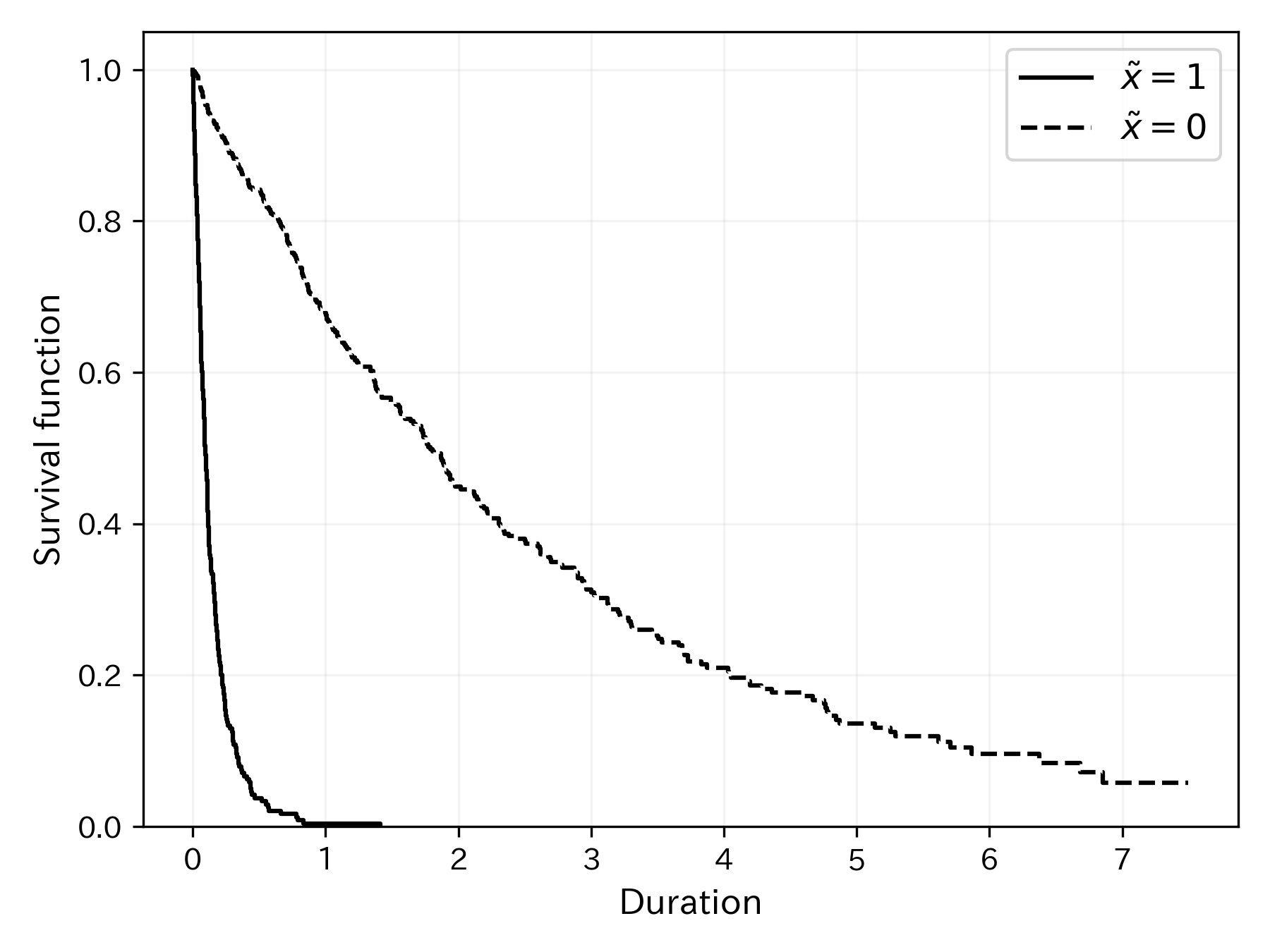}%
    }\hspace{1em}
    \subcaptionbox{Kaplan–Meier Estimator\label{fig:scenario-b-stratified-km}}{%
        \includegraphics[width=0.48\linewidth, height=0.22\textheight, keepaspectratio]{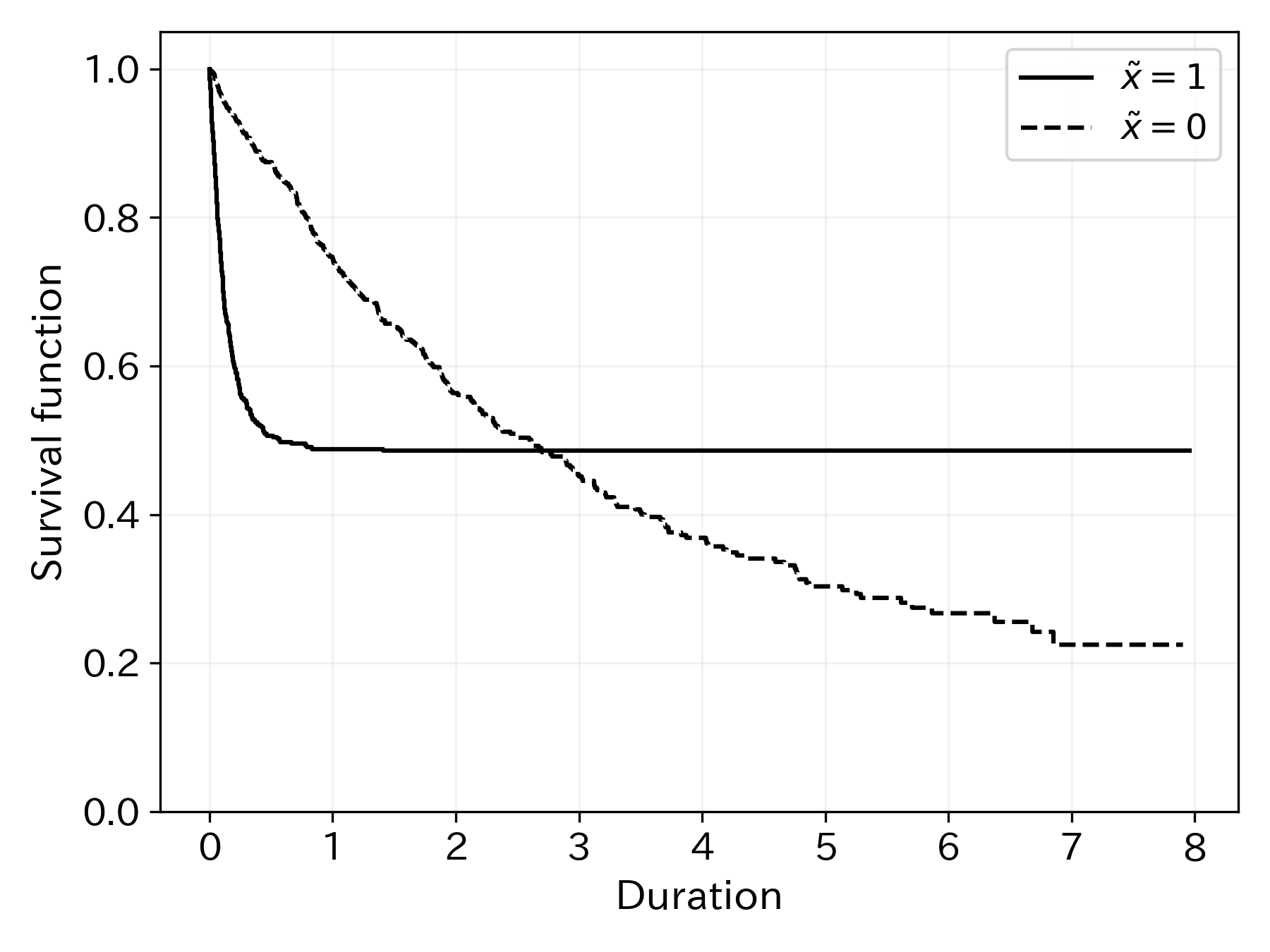}%
    }

    \subcaptionbox{Incidence Coefficient Estimates\label{fig:scenario-b-incidence-coef}}{%
        \includegraphics[width=0.48\linewidth, height=0.22\textheight, keepaspectratio]{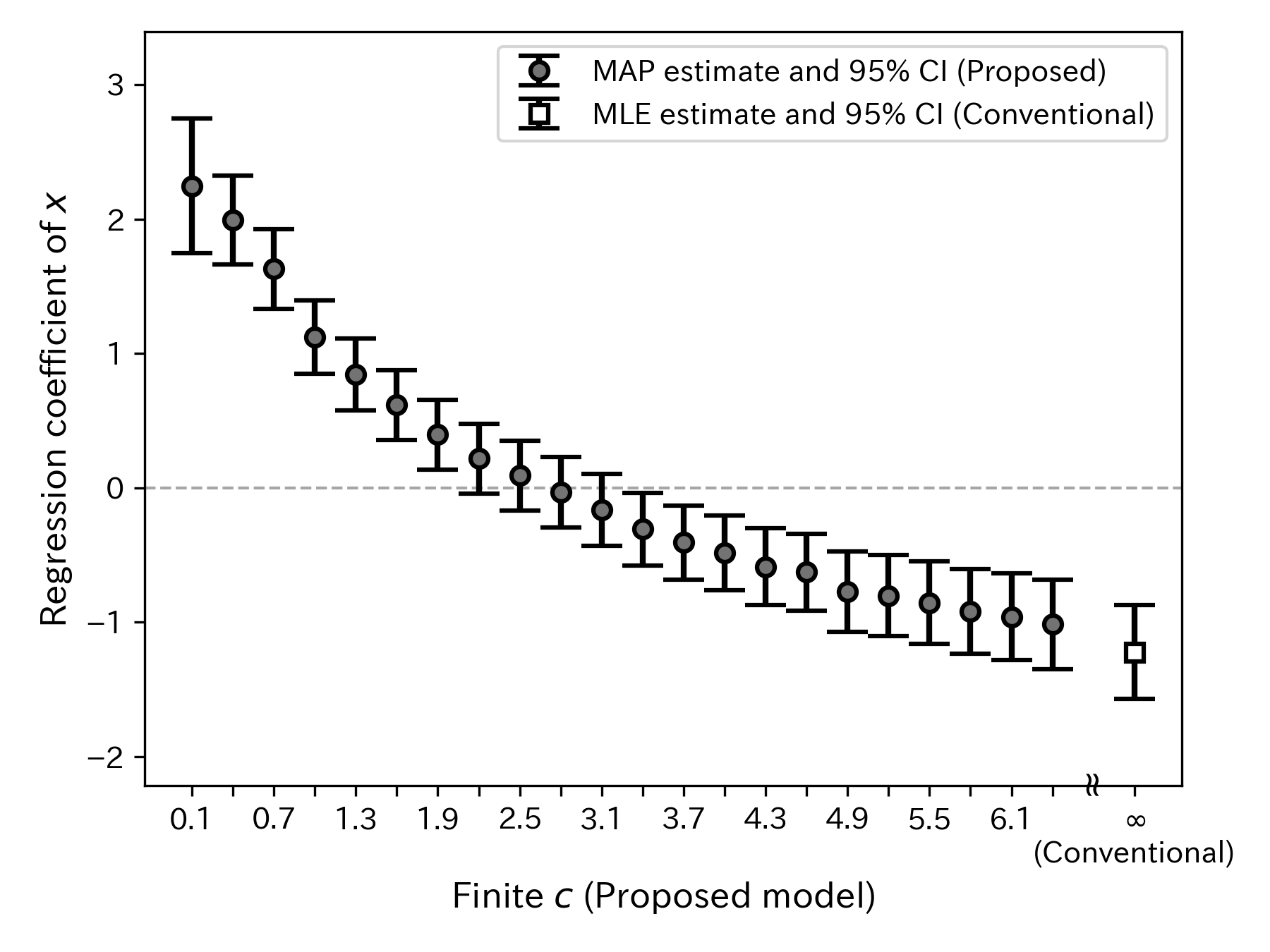}%
    }\hspace{1em}
    \subcaptionbox{Latency Coefficient Estimates\label{fig:scenario-b-latency-coef}}{%
        \includegraphics[width=0.48\linewidth, height=0.22\textheight, keepaspectratio]{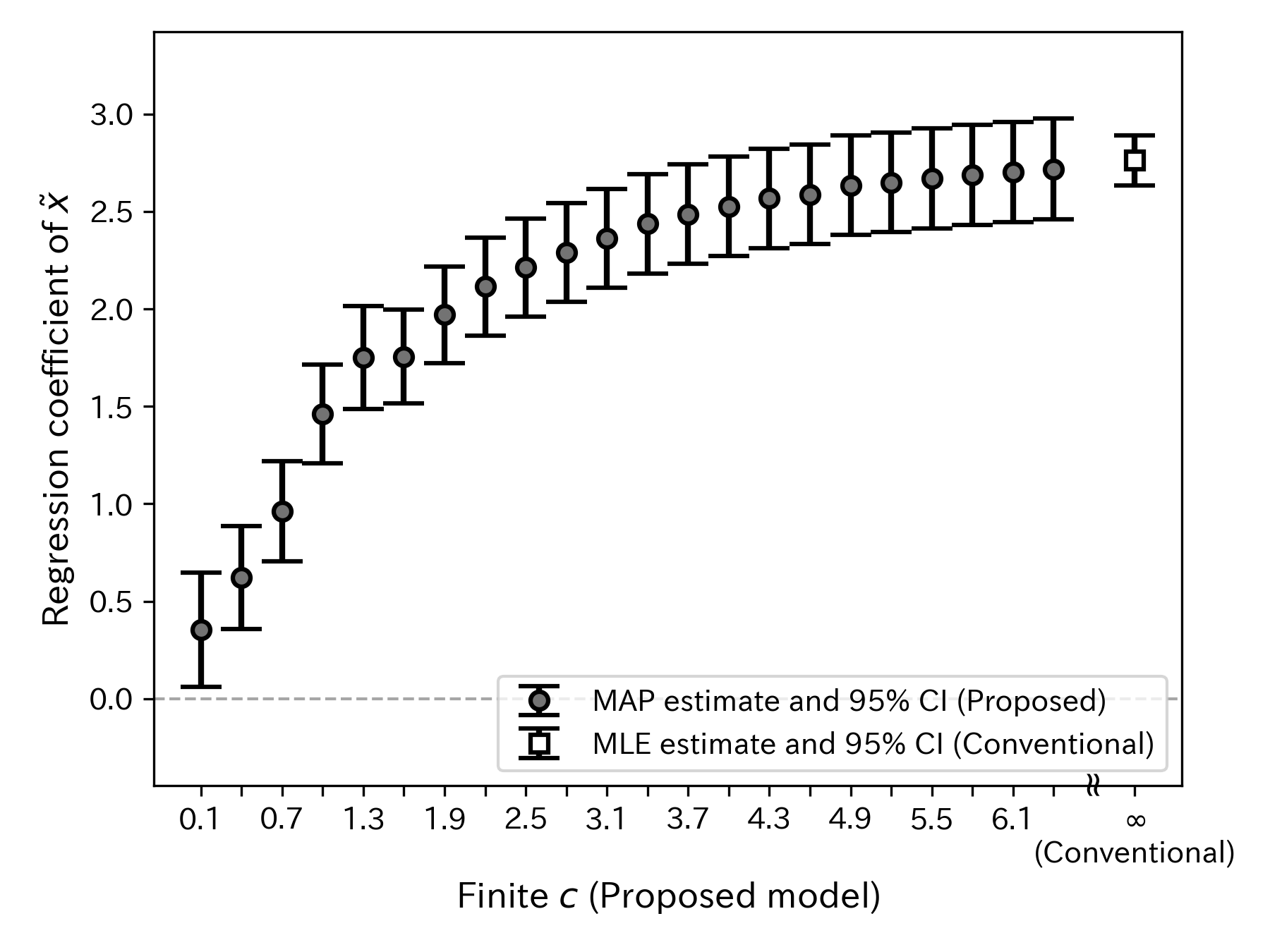}%
    }

    \caption{Summary of results for Scenario B.}
    \label{fig:scenario-b}
\end{figure}

In the simulation for this scenario, we consider a binary covariate $x$ (e.g., treatment vs.\ control, or exposed vs.\ unexposed). The covariate is generated as $x \sim \mathrm{Bernoulli}(0.5)$. Conditional on $x$, the ultimate event indicator $Y \in \{0,1\}$ (whether the event ever occurs) and the event time $T$ are generated from a cure-type exponential model. The incidence part is specified so that
\begin{equation*}
    \pi(x) = P(Y=1 \mid x) =
    \begin{cases}
        0.5, & \text{if } x = 1, \\
        0.8, & \text{if } x = 0,
    \end{cases}
\end{equation*}
and, given $Y=1$, the latency part follows an exponential distribution
\begin{equation*}
    T \mid (Y=1, x) \sim \mathrm{Exp}(\lambda(x)),
    \qquad
    \lambda(x) =
    \begin{cases}
        7.0, & \text{if } x = 1, \\
        0.4, & \text{if } x = 0.
    \end{cases}
\end{equation*}
Thus $x=1$ reduces the ultimate probability of the event but shortens the time to the event among susceptible individuals. Figure~\ref{fig:scenario-b-only-susceptible-km} shows the Kaplan–Meier estimator for the susceptible-only population. The susceptible-only curve for $x=1$ approaches zero within the first few time units. Figure~\ref{fig:scenario-b-stratified-km} shows the Kaplan–Meier estimator for the overall population. The $x=1$ group exhibits an early concentration of events, with the survival curve dropping sharply and then flattening around 0.5, whereas the $x=0$ group shows a more gradual decline. Independent censoring times are sampled as $C \sim \mathrm{Uniform}(0,8)$, and we observe $(t_i,\delta_i) = (\min\{T_i,C_i\},\mathbf{1}\{T_i \le C_i\})$ for $n = 1{,}000$ individuals.

When we fit the conventional mixture cure model \citep{2000Sy_EstimationCoxProportional} that assumes an infinite time horizon, using $x$ as the covariate in both the incidence and latency components, the estimated incidence coefficient for $x$ is negative, whereas the latency coefficient is positive (Figures~\ref{fig:scenario-b-incidence-coef} and \ref{fig:scenario-b-latency-coef}). Thus the conventional model suggests that $x=1$ reduces the eventual probability of the event, which is consistent with the data-generating mechanism ($\pi(1) < \pi(0)$). In contrast, when we fit the proposed finite-horizon mixture model with various cutoffs $c \in \{0.1, 0.5, \ldots, 6.1\}$, the estimated coefficients for $x$ in the incidence change sign as $c$ increases. This captures that $x=1$ increases the probability of the event within the $c$-period and shortens the time to the event. For larger $c$, the incidence coefficient decreases and eventually turns negative, reflecting the transition from the short-horizon question (event probability within $c$) to the long-horizon question (ultimate event probability). This means the effect of $x$ on $\mathbb{P}(T < c \mid x=1)$ changes as $c$ changes. If we want to know the effect of $x$ on $\mathbb{P}(T < c \mid x=1)$ for a specific $c$, we need to fit the model with the cutoff $c$.
This sign reversal illustrates how the choice of time horizon $c$ fundamentally affects the interpretation of the incidence effect. In cases like the present data, where this tendency is clearly observed, it may be possible to detect it. However, in real-world data, many variables are involved, so this tendency may not be identifiable.

\section{Real Data Analysis}\label{sec:real-data-analysis}

In this section, we apply the proposed model to transaction data from Mercari and analyze user behavior in the online flea market.

\subsection{Data Description}

Mercari is an online flea market platform operated by Mercari, Inc., where users can list items for sale and other users can purchase those items. The Mercari dataset contains structured information such as item prices, categories, and shipping methods, as well as unstructured information including item titles, descriptions, and thumbnail images. Seller-specific information, such as transaction history or reputation measures, is not available. We analyze the time from product listing to transaction completion as survival time. A transaction completion involves a sequence of steps, including payment by the buyer, shipment by the seller, receipt of the item, and submission of a review. In this study, a transaction completion is treated as the event of interest ($\delta_{i}$ = 1). Items that have not completed a transaction by the data extraction date are treated as right-censored ($\delta_{i}$ = 0). The target population consists of items listed in 2020, and the dataset is a snapshot collected in June 2023. As a result, temporal changes in product attributes after listing (e.g., price updates, description edits, or image changes) are not observed. Items that remain unsold as of June 2023 are therefore censored. Under this setup, it is possible to observe whether an item was sold for at least two and a half years after listing.

\begin{figure}
    \centering
    \includegraphics[width=\textwidth,height=0.4\textheight,keepaspectratio]{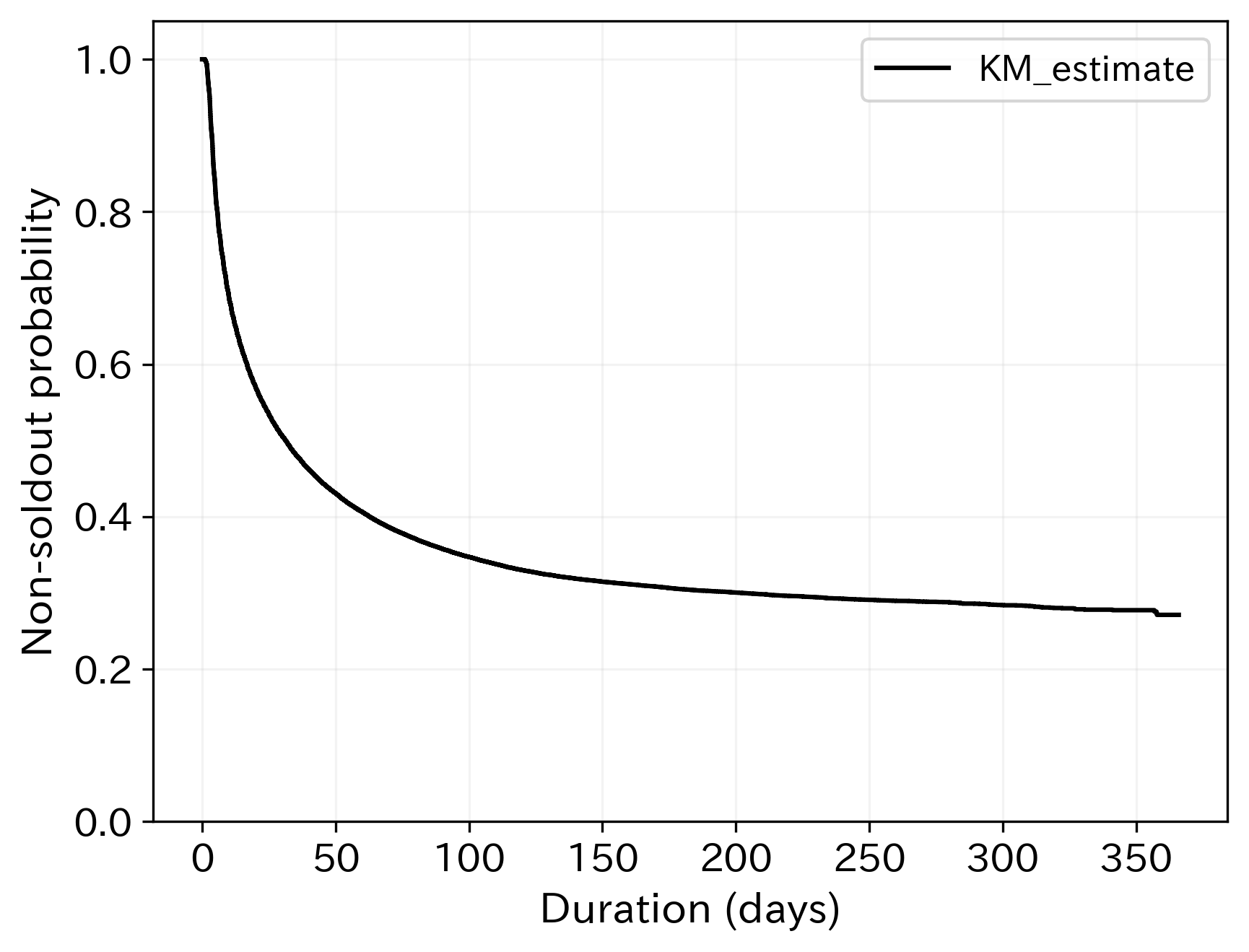}
    \caption{Kaplan–Meier Estimator}
    \label{fig:km_plot_tshirt}
\end{figure}

Since the dataset spans from 2020 to June 2023, every item listed in 2020 has at least two and a half years of follow-up, and whether a transaction occurred within any reasonable horizon $c$ is directly observed. To demonstrate the proposed model in the more realistic setting where some individuals are right-censored before $c$, we introduce an artificial administrative censoring date of January 1, 2021, assuming that product attributes observed in June 2023 are identical to those at that date. Items unsold by January 1, 2021 are treated as censored at that date. In this censoring mechanism, the survival and censoring times can be considered independent conditional on covariates. Mercari offers a wide range of product categories, including men's, women's, tickets, and furniture. In this analysis, we focus on items belonging to the most granular category, T-shirts/Cut and Sewn (Short Sleeve/Sleeveless), while retaining higher-level categories (e.g., men's or women's) as covariates. Among highly frequent fine-grained categories—such as "Others," "Idol," "Manga," and T-shirts/Cut and Sewn (Short Sleeve/Sleeveless)—we selected this category under the assumption that attribute variability within the category is relatively small. From this category, we randomly sampled 100,000 items, restricting attention to items priced below 15,000 JPY and with a survival time of at least one day. This restriction is motivated by the fact that transaction completion includes shipping after purchase, which we assume takes at least one day; survival times shorter than one day are therefore treated as outliers and excluded. After this filtering, the sample size is $N = 98{,}258$. Descriptive statistics are summarized in Table~\ref{tab:mercari_data_description}. As shown in the table, the dataset contains many categorical variables, with price being the only continuous variable. The number of listings increases from March, peaks in May, and then decreases monotonically toward December, indicating seasonality in listing activity. The standard deviation of prices is also relatively large. Figure~\ref{fig:km_plot_tshirt} presents the Kaplan–Meier estimate of the survival function for the sampled data. The survival curve drops sharply immediately after listing and then levels off around 30\%, suggesting that most items are sold shortly after listing, while a substantial fraction remains unsold. It is unlikely that all unsold items would eventually complete a transaction given a sufficiently long listing period, which motivates the use of cure-type models. Beyond the generic case for cure-type modeling, the C2C holding-cost structure discussed in Section~\ref{sec:introduction} applies directly to Mercari: sellers are individual consumers whose welfare depends on selling within a tolerable period, so the substantive estimand is $\mathbb{P}(T < c \mid x)$ for a finite $c$ rather than $\mathbb{P}(T < \infty \mid x)$.
In line with this rationale, we set the cutoff at $c = 1$ week and $3$ months in the proposed model, representing horizons within which the seller can plausibly tolerate holding the item. Under this choice of $c$, the proposed model can be interpreted as evaluating (i) the effect of covariates on the probability of transaction completion within one week or three months, respectively, and (ii) the effect of covariates on the speed of transaction completion among items that complete a transaction within this period.

\begin{table}[htbp]
    \centering
    \caption{Descriptive statistics of the Mercari data (T-shirts/Cut and Sewn (Short Sleeve/Sleeveless))}
    \label{tab:mercari_data_description}
    \renewcommand{\arraystretch}{1.2}
    \begin{tabular}{p{0.40\linewidth}p{0.55\linewidth}}
        \toprule
        \multicolumn{2}{l}{\textbf{Sample Composition}}                                                                                                                                                         \\
        \midrule
        Sample size $N$                & 98,258                                                                                                                                                                 \\
        Number of observed events      & 66,298                                                                                                                                                                 \\
        Number of censored             & 31,960                                                                                                                                                                 \\
        Censoring rate                 & 32.5\%                                                                                                                                                                 \\
        Observed survival time (hours) & min 25 / Q1 169 / median 675 / Q3 2,937 / max 8,783                                                                                                                    \\
        \midrule
        \multicolumn{2}{l}{\textbf{Main Attributes}}                                                                                                                                                            \\
        \midrule
        Top-level category             & Men's 55.2\%, Women's 44.8\%                                                                                                                                           \\
        Item condition                 &
        No noticeable damage or stains 43.4\%,
        Brand new/unused 30.2\%,
        Almost unused 13.3\%,
        Slightly damaged/stained 10.7\%,
        Damaged/stained 2.1\%,
        Overall poor condition 0.3\%                                                                                                                                                                            \\
        Shipping lead time             & 1--2 days 51.4\%, 2--3 days 34.5\%, 4--7 days 14.1\%                                                                                                                   \\
        Shipping charge burden         & Shipping included (seller) 99.0\%, Cash on delivery (buyer) 0.97\%                                                                                                     \\
        Listing month                  & Jan 3.0\%, Feb 3.5\%, Mar 6.3\%, Apr 10.0\%, May 19.2\%, Jun 14.4\%, Jul 13.6\%, Aug 12.5\%, Sep 7.2\%, Oct 4.4\%, Nov 3.1\%, Dec 2.6\%                                \\
        Anonymous shipping             & Anonymous 86.5\%, Non-anonymous 13.5\%                                                                                                                                 \\
        Price                          & min 300 JPY, Q1 800 JPY, median 1,500 JPY, Q3 2,800 JPY, max 15,000 JPY, mean 2,266.3 JPY (SD 2,342.3 JPY)                                                             \\
        Size                           & M 36.7\%, L 22.3\%, S 14.9\%, FREE SIZE 12.4\%, XL(LL) 8.8\%, XS(SS) 2.1\%, 2XL(3L) 1.8\%, 3XL(4L) 0.6\%, 4XL(5L) or larger 0.3\%, XXS or smaller 0.2\%, missing 0.1\% \\
        \bottomrule
    \end{tabular}
\end{table}

\subsection{Data Preprocessing and Model Fitting}

For the application of the analytical models, we conducted data preprocessing. Specifically, the dataset was randomly split into training and test sets with a 4:1 ratio. We then prepared covariates by creating dummy variables for whether the listing date was on a holiday or weekend and for each month, as well as for categorical features such as brand, size, item condition, shipping lead time, and shipping charge burden. For brands, we extracted the top five most frequent brands in both the men's and women's training data, and created dummy variables for their union (excluding 'No brand'). For each categorical predictor (including listing month), the reference category was set to the most frequent level in the training data. Variables that may be modified after listing (item name, and item description) were excluded from the analysis to focus on intrinsic product attributes and listing conditions that are fixed at the time of listing. Finally, an intercept was added and covariates for each item $x_{i}$ were obtained, resulting in $p=41$. In addition to the proposed model, we used the conventional model \citep{2000Sy_EstimationCoxProportional} as a benchmark for analysis. Training was performed on the training set, and evaluation was carried out on the test set.

The main hyperparameter for the proposed model, $K$, was chosen by comparing the training-data posterior $p(\theta \mid \mathcal{D})$ across $K \in \{10, 20, 30, 40\}$; gains from increasing $K$ beyond $20$ were negligible, so we fixed $K=20$. A sensitivity analysis with respect to $K$ indicated that the substantive results were largely unchanged. For cross-model comparison, we measured $\overline{\mathrm{AUC}}(25,c)$ \citep{2020Polsterl_scikit-survival,2007Uno_EvaluatingPredictionRules} on the test data at $c = 1$ week and $c = 3$ months (Table~\ref{tab:auc_horizons}), with all times in hours so that the lower endpoint $25$ coincides with the minimum observed duration after excluding listings with time-to-sale of at most one day. Because both models use the same covariate structure and differ only in the target of inference (finite vs.\ infinite horizon), differences in $\overline{\mathrm{AUC}}(25,c)$ are expected to be small; the present study does not deliberately alter model flexibility.

As for model interpretation, we compare the effects of gender, size, and listing month between the proposed model ($c = 1$ week and $c = 3$ months) and the conventional model; among the covariates examined, these three showed the largest discrepancies across the two approaches.

\begin{figure}
    \centering
    \includegraphics[width=0.8\textwidth,height=0.4\textheight,keepaspectratio]{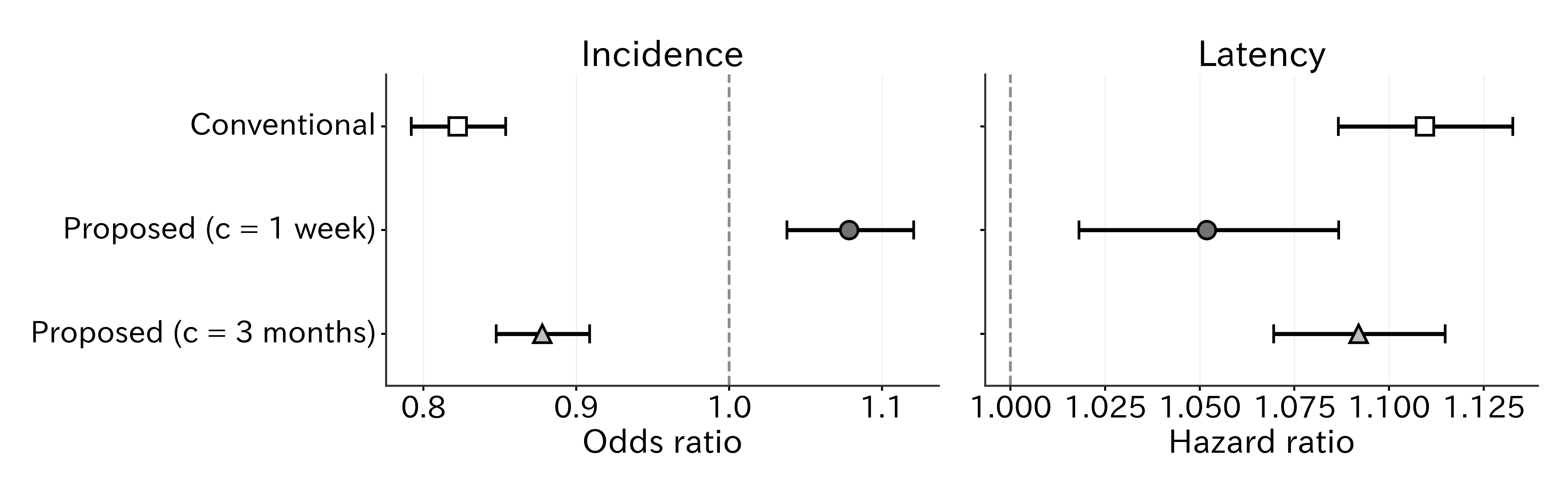}
    \caption{Comparison of the estimated Women's effect with Men's as the reference category.}
    \label{fig:gender_proposed_vs_previous}
\end{figure}

Figure~\ref{fig:gender_proposed_vs_previous} compares the Women's effect with Men's as the reference category. The conventional mixture cure model decomposes covariate effects into an incidence component for the odds of eventual susceptibility ($T < \infty$) and a latency component for the hazard of sale timing conditional on $T < \infty$, whereas the proposed model targets the odds of sale within horizon $c$ and the hazard of sale timing conditional on $T < c$ among listings with $T < c$. Under $c = 1$ week, the odds ratio exceeds one, so Women's listings are estimated to have a higher probability than Men's of selling within one week. By contrast, for the proposed model with $c = 3$ months and for the conventional model, the odds ratios are below one and the hazard ratios are above one, so the directions of the estimated odds and hazard effects agree between these two specifications. Although the estimated coefficients share the same signs across these two specifications, the underlying model structures differ, so these coefficients do not admit the same interpretation.

\begin{figure}
    \centering
    \begin{subfigure}{\textwidth}
        \centering
        \includegraphics[width=\textwidth,height=0.38\textheight,keepaspectratio]{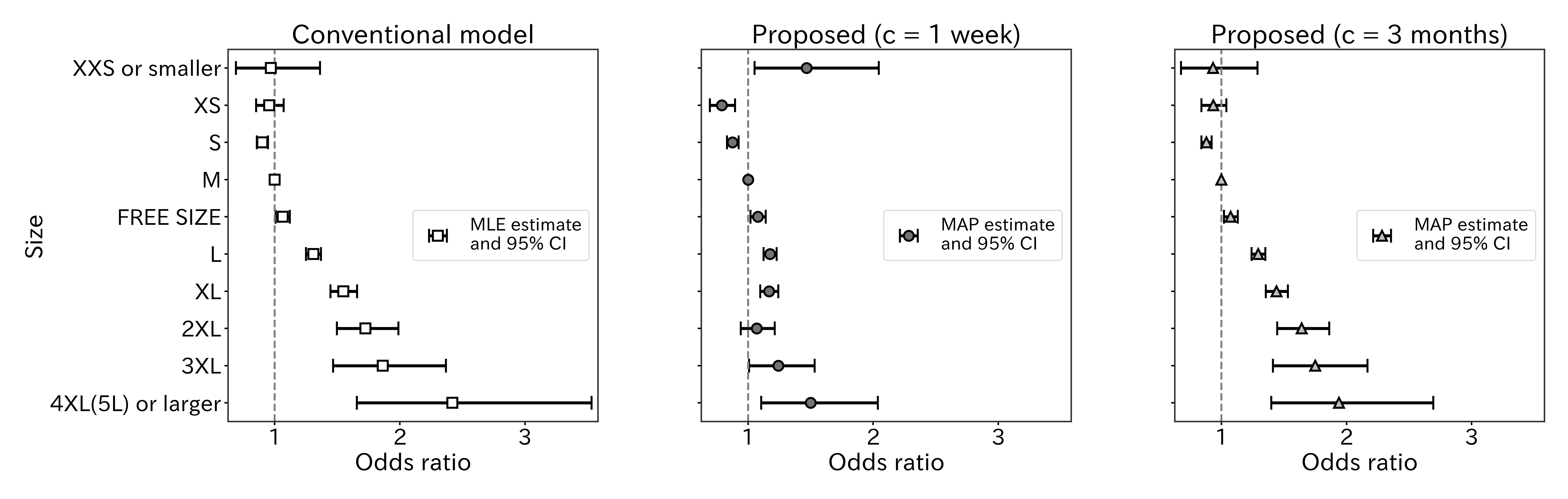}
        \caption{Odds ratio.}
        \label{fig:size_proposed_vs_conventional_odds}
    \end{subfigure}

    \medskip

    \begin{subfigure}{\textwidth}
        \centering
        \includegraphics[width=\textwidth,height=0.38\textheight,keepaspectratio]{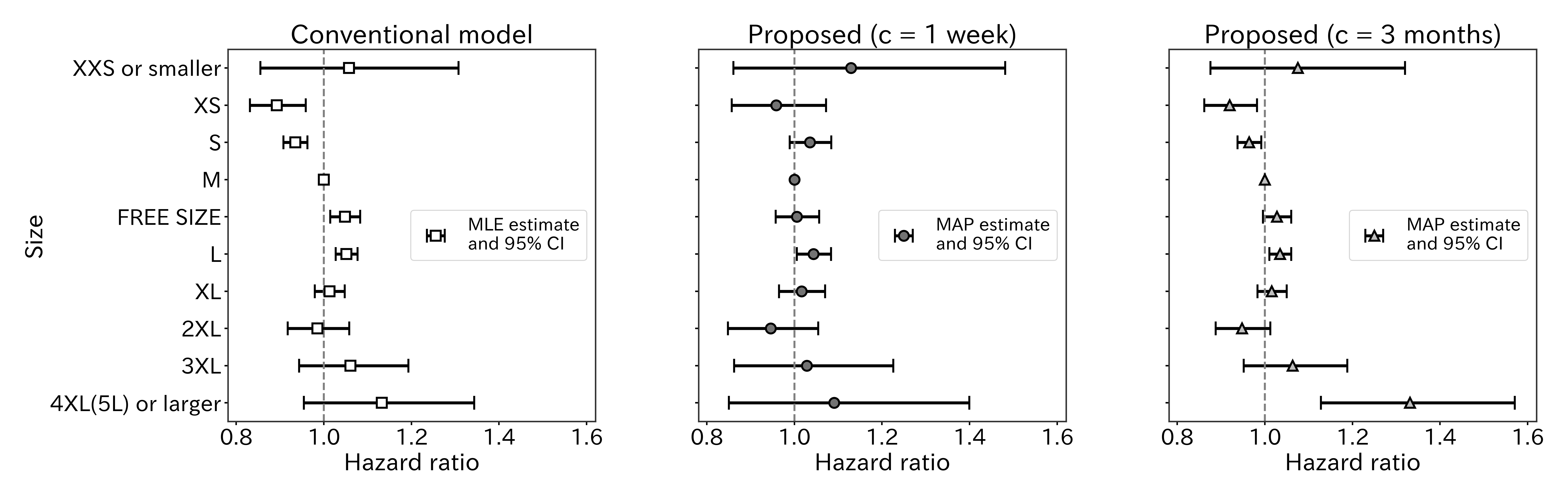}
        \caption{Hazard ratio.}
        \label{fig:size_proposed_vs_conventional_hazard}
    \end{subfigure}
    \caption{Comparison of size effect with M as the reference category.}
    \label{fig:size_proposed_vs_conventional}
\end{figure}

Figure~\ref{fig:size_proposed_vs_conventional} compares size effects with M as the reference category (Figure~\ref{fig:size_proposed_vs_conventional_odds} for odds ratios; Figure~\ref{fig:size_proposed_vs_conventional_hazard} for hazard ratios). For odds, the proposed model with $c = 1$ week indicates that XXS or smaller has a significantly positive effect on the probability of completing a transaction within one week. By contrast, for the conventional model---which targets the odds of eventually completing a transaction---and for the proposed model with $c = 3$ months---which targets the odds of completing a transaction within three months---XXS or smaller is not significant. At the same time, 4XL(5L) or larger is significantly positive for odds in all three specifications. For hazard ratios, only the proposed model with $c = 3$ months assigns a significantly positive effect to 4XL(5L) or larger. Taken together, these estimates suggest that on Mercari, extremely small and extremely large sizes---sizes that are rarely offered in mainstream business-to-consumer apparel retail---materially affect both transaction completion and how readily a completion occurs within the relevant horizon. For odds ratios and for hazard ratios separately, the covariates that emerge as statistically significant can differ depending on whether the conventional model or the proposed model with a given horizon $c$ is used; care is therefore required when translating significance patterns into decisions. In particular, if the operational objective is to complete a transaction within one week but the conventional model is used, the significant covariates can differ from those under the proposed model with $c = 1$ week, which may lead to erroneous decisions.

\begin{figure}
    \centering
    \begin{subfigure}{\textwidth}
        \centering
        \includegraphics[width=\textwidth,height=0.38\textheight,keepaspectratio]{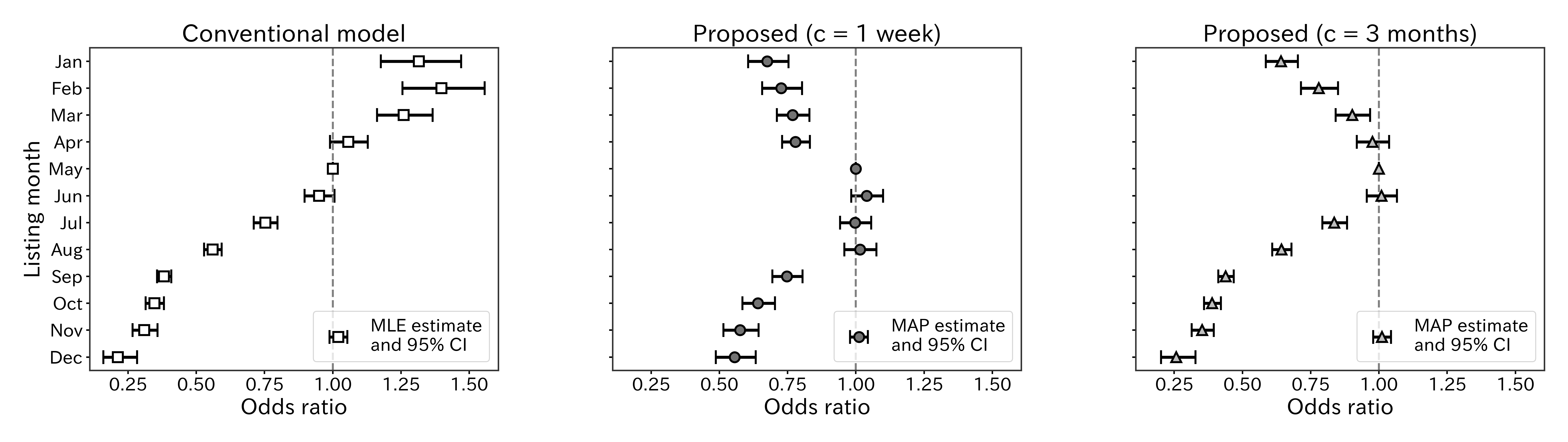}
        \caption{Odds ratio.}
        \label{fig:listing_month_proposed_vs_conventional_odds}
    \end{subfigure}

    \medskip

    \begin{subfigure}{\textwidth}
        \centering
        \includegraphics[width=\textwidth,height=0.38\textheight,keepaspectratio]{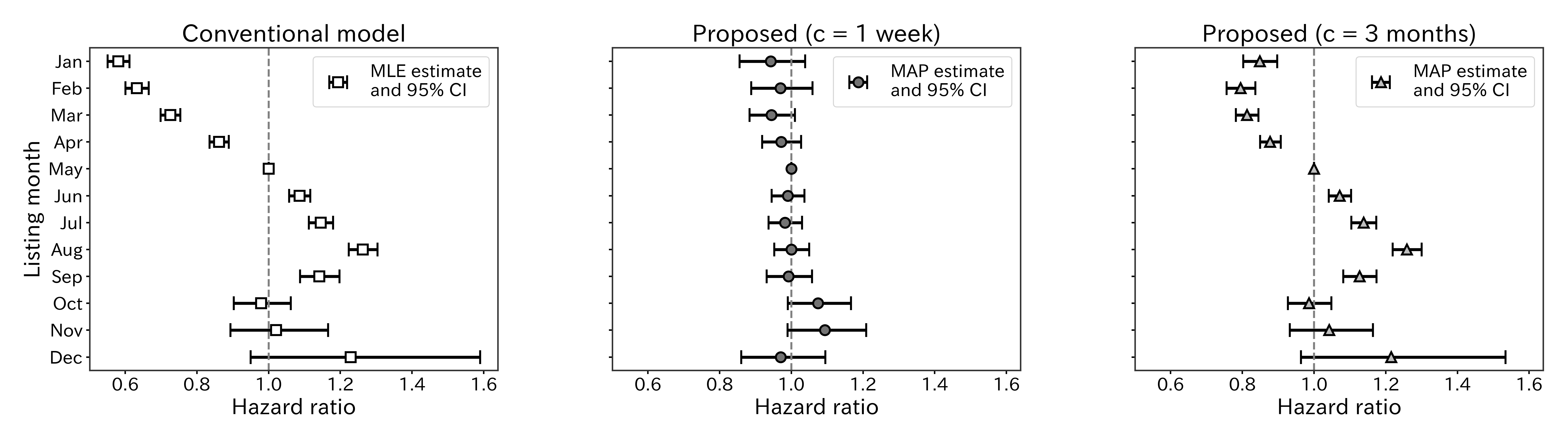}
        \caption{Hazard ratio.}
        \label{fig:listing_month_proposed_vs_conventional_hazard}
    \end{subfigure}
    \caption{Comparison of listing month effect with May as the reference category.}
    \label{fig:listing_month_proposed_vs_conventional}
\end{figure}

Figure~\ref{fig:listing_month_proposed_vs_conventional} summarizes listing-month effects with May as the reference month: Figure~\ref{fig:listing_month_proposed_vs_conventional_odds} reports odds ratios and Figure~\ref{fig:listing_month_proposed_vs_conventional_hazard} reports hazard ratios. The interpretive gap between the conventional and proposed specifications is largest for the odds ratios. In the proposed model with $c = 1$ week, the coefficient for January contrasts the odds that an item listed in January completes a transaction within one week of listing relative to items listed in May. In the proposed model with $c = 3$ months, the coefficient for January contrasts the odds of completing a transaction within three months of listing---for items listed in January, this corresponds to sale by approximately April relative to the May benchmark. The conventional model shows relatively large odds ratios in January and February, followed by a monotone decline toward December. By contrast, the proposed model with $c = 1$ week yields the strongest odds ratios from May through August relative to the other months, whereas the proposed model with $c = 3$ months shows an increase from January to May and a decline from July onward. Because the data analyzed here are for T-shirts/Cut and Sewn (Short Sleeve/Sleeveless) items, a pattern reflecting stronger demand in months closer to the summer season in Japan is substantively plausible. This seasonal pattern is not mirrored by the conventional odds profile, whereas the proposed odds profiles are more consistent with such domain knowledge. For hazard ratios, the conventional model and the proposed model with $c = 3$ months both exhibit seasonality in listing month, whereas the proposed model with $c = 1$ week yields estimates that are comparatively flat across months, suggesting little seasonality in the hazard among items that sell within one week. A plausible interpretation is that listings that clear within one week are broadly popular items for which seasonal variation is secondary. Aligning the estimand---and hence the model and horizon $c$---with the decision criterion therefore facilitates interpretations that match the operational objective.

Finally, regarding the nearly monotone decline of the conventional model's odds ratios for listing month in the odds-ratio panel (Figure~\ref{fig:listing_month_proposed_vs_conventional_odds}), we considered whether the observation period might affect the estimation of the coefficients. For example, items listed in January can be observed for at least 330 days, while those listed in December can be observed for at most 30 days. To examine this issue, we also conducted the same conventional model estimation for categories other than T-shirts/Cut and Sewn (Short Sleeve/Sleeveless). As a result, the estimated coefficients did not show a monotonic decrease, and thus we concluded that the observation period in the conventional model does not have a major impact on the estimation of the coefficients.

\begin{table}
    \centering
    \caption{$\overline{\mathrm{AUC}}(25,c)$ on the test data by horizon $c$.}
    \label{tab:auc_horizons}
    \begin{tabular}{lcc}
        \toprule
        Model              & $c = 1$ week & $c = 3$ months \\
        \midrule
        Conventional model & 0.5841       & 0.5979         \\
        Proposed model     & 0.5960       & 0.5995         \\
        \bottomrule
    \end{tabular}
\end{table}

\section{Conclusion}\label{sec:conclusion}

This study proposed a mixture cure model based on a prespecified finite time $c$, in which the population is latently divided according to whether the event occurs within $[0, c)$ rather than at infinite time. By imposing the boundary constraint $\lim_{t \to c-} S_c(t \mid x) = 0$, the model achieves identifiability without relying on untestable assumptions about tail behavior, and classifies censored observations with $t_i \ge c$ deterministically as belonging to the event-free group, thereby extracting
additional information relative to the conventional formulation.

Simulation studies verified the statistical properties of the proposed estimator.  In Scenario A, empirical bias and standard deviation decreased as the sample size increased, and coverage probabilities remained close to the nominal 95\% level across a range of event rates and covariate structures, indicating that the EM-based estimator behaves as expected.  Scenario B revealed a qualitatively important pitfall of conventional infinite-horizon models: when the covariate simultaneously increases the short-term event
probability and reduces the long-term probability, the incidence coefficient estimated by the conventional model can reverse sign relative to the finite-horizon truth. This sign reversal demonstrates that using an
infinite-horizon model for finite-horizon decision-making can lead to erroneous conclusions, and it underscores the practical value of explicitly specifying a decision-relevant horizon $c$.

In the application to the Mercari dataset, the proposed model---fit with a three-month horizon reflecting the operational decision window---identified qualitatively different patterns from the conventional model.  The estimated listing month effect captured a summer-season demand peak consistent with prior knowledge about short-sleeve apparel, a pattern that was not apparent in the conventional model.  The significance of size effects also differed between the two models, illustrating that the choice of estimand (finite
vs.\ infinite horizon) can alter substantive conclusions in practice.

A key characteristic of the proposed framework is that all model interpretations are explicitly scoped to the analyst-specified horizon $c$, which is chosen a priori to reflect the decision-relevant time window rather than inferred from data. Accordingly, the model makes no claim about event probabilities beyond $c$, and conclusions may differ across different choices of $c$, as demonstrated in Scenario B. When the appropriate horizon is unclear, sensitivity analysis over multiple values of $c$ is advisable. Future extensions include the incorporation of time-dependent covariates, left truncation, interval censoring, and the derivation of asymptotic distributional theory for the proposed estimators.

\section*{Acknowledgements}
This study was supported by the Sunaga Shigemitsu Economics Support Fund of the Tohoku University Fund. We also used the Mercari Dataset provided by Mercari, Inc. through the IDR Dataset Provision Service of the National Institute of Informatics. We hereby express our gratitude to all those involved.

\appendix

\section{Empirical Bayes Method}\label{sec:empirical-bayes-method}
We want to find the hyperparameter $\lambda$ that maximizes:
\begin{align}
    p(\mathcal{D}|\lambda) = \int p(\mathcal{D}\mid \theta) p(\theta | \lambda) d\theta
    \label{marginal_likelihood}
\end{align}
Here, let the dimensions of $\theta, b, \beta, \alpha$ be $M, M_{b}, M_{\beta}, M_{\alpha}$, respectively. We consider this following \cite{MacKay1992BayesianInterpolation}. In many cases, $p(\mathcal{D}|\lambda)$ cannot be obtained analytically. Therefore, letting
\begin{align*}
    f(\theta) = p(\mathcal{D}\mid \theta) p(\theta | \lambda)
\end{align*}
we have
\begin{align*}
    \log{f(\theta)} = \log{p(\mathcal{D}\mid \theta)} + \log{p(\theta | \lambda)}
\end{align*}
Considering MAP estimation,
\begin{align*}
    \hat{\theta} = \operatorname*{argmax}_{\theta} f(\theta)
\end{align*}
Since $\nabla_{\theta}f(\theta)|_{\theta=\hat{\theta}} = 0$, using the Laplace approximation yields:
\begin{align*}
    f(\theta) \approx f(\hat{\theta}) \exp{\left(-\frac{1}{2}(\theta - \hat{\theta})^\top A (\theta - \hat{\theta})\right)}
\end{align*}
where $A = - \nabla^{2}_{\theta}\log{f(\theta)}|_{\theta=\hat{\theta}}$. Integrating this with respect to $\theta$,
\begin{align*}
    p(\mathcal{D}|\lambda) \approx f(\hat{\theta}) (2\pi)^{\frac{M}{2}} |A|^{-\frac{1}{2}}
\end{align*}
Taking the logarithm,
\begin{align*}
    \log{p(\mathcal{D}|\lambda)} \approx \log{f(\hat{\theta})} + \frac{M}{2} \log{2\pi}-\frac{1}{2} \log{|A|}
\end{align*}
From $f(\hat{\theta}) = p(\mathcal{D}|\hat{\theta})p(\hat{\theta}|\lambda)$, we obtain:
\begin{align}
    \log{p(\mathcal{D}|\lambda)} \approx \log{p(\mathcal{D}|\hat{\theta})} + \log{p(\hat{\theta}|\lambda)} + \frac{M}{2} \log{2\pi}-\frac{1}{2} \log{|A|}
    \label{eq:laplace_log_marginal_likelihood}
\end{align}
Here, if we set $p(\theta \mid \lambda) = p(b \mid \lambda)p(\beta \mid \lambda)p(\alpha \mid \lambda)\propto p(\alpha \mid \lambda)$, then $p(\alpha \mid \lambda) = \mathcal{N}(\alpha|0, \lambda^{-1} I_{M_{\alpha}}) = \left(\frac{\lambda}{2\pi}\right)^{\frac{M_{\alpha}}{2}} \exp{\left(-\frac{\lambda}{2} \alpha^\top \alpha\right)}$. Taking the logarithm gives:
\begin{align}
    \log{p(\alpha|\lambda)} & = \frac{M_{\alpha}}{2}\log{\lambda} - \frac{M_{\alpha}}{2}\log{2\pi} - \frac{\lambda}{2} \alpha^\top \alpha \nonumber \\
                            & = \frac{M_{\alpha}}{2} \log{\lambda} - \frac{\lambda}{2} \alpha^\top \alpha + \mathrm{const}
    \label{eq:log_gaussian_prior_lambda}
\end{align}
Substituting Equation \eqref{eq:log_gaussian_prior_lambda} into Equation \eqref{eq:laplace_log_marginal_likelihood}, we get:
\begin{align}
    \log{p(\mathcal{D}|\lambda)} \approx \log{p(\mathcal{D} | \hat{\theta})} + \frac{M_{\alpha}}{2} \log{\lambda} - \frac{\lambda}{2} \hat{\alpha}^\top \hat{\alpha} - \frac{1}{2} \log{|A|} + \mathrm{const}
    \label{eq:laplace_log_marginal_likelihood_lambda}
\end{align}
Here,
\begin{align*}
    A & = - \nabla^{2}_{\theta}\log{f(\theta)}|_{\theta=\hat{\theta}}                                                                                 \\
      & = - \nabla^{2}_{\theta}\log{p(\mathcal{D}|\theta)}|_{\theta=\hat{\theta}} - \nabla^{2}_{\theta}\log{p(\theta|\lambda)}|_{\theta=\hat{\theta}} \\
      & = H + K
\end{align*}
where $H = - \nabla^{2}_{\theta}\log{p(\mathcal{D}|\theta)}|_{\theta=\hat{\theta}}$ and $K = - \nabla^{2}_{\theta}\log{p(\theta|\lambda)}|_{\theta=\hat{\theta}}$.
Furthermore, letting $\tilde{\theta} = (b, \beta)$, with $\theta = (\tilde{\theta}, \alpha)$, we define:
\begin{align*}
    H =
    \begin{pmatrix}
        H_{\tilde\theta\tilde\theta} & H_{\tilde\theta\alpha} \\
        H_{\alpha\tilde\theta}       & H_{\alpha\alpha}
    \end{pmatrix},
    \quad
    H_{xy}=-\nabla^2_{xy}\log p(\mathcal D\mid\theta)\big|_{\theta = \hat\theta}.
\end{align*}
Then $A$ can be written as:
\begin{align*}
    A = \begin{pmatrix}
            H_{\tilde\theta\tilde\theta} & H_{\tilde\theta\alpha}                    \\
            H_{\alpha\tilde\theta}       & H_{\alpha\alpha} + \lambda I_{M_{\alpha}}
        \end{pmatrix}
\end{align*}
By decomposition using the Schur complement,
\begin{align*}
    |A|
     & = |H_{\tilde\theta\tilde\theta}| \cdot \left| H_{\alpha\alpha} + \lambda I_{M_{\alpha}} - H_{\alpha\tilde\theta} H_{\tilde\theta\tilde\theta}^{-1} H_{\tilde\theta\alpha} \right|                              \\
     & = |H_{\tilde\theta\tilde\theta}| \cdot \left| S + \lambda I_{M_{\alpha}} \right|  \quad \left( S := H_{\alpha\alpha} - H_{\alpha\tilde\theta} H_{\tilde\theta\tilde\theta}^{-1} H_{\tilde\theta\alpha} \right)
\end{align*}
Here, since $S$ is a real symmetric matrix, it is diagonalizable by an orthogonal matrix. Let its eigenvalues be $\mu_{1}, \ldots, \mu_{M_{\alpha}}$. Letting $v_{i}$ be the eigenvector corresponding to $\mu_{i}$,
\begin{align*}
    S v_{i} = \mu_{i} v_{i}
\end{align*}
\begin{align*}
    (S + \lambda I_{M_{\alpha}})v_{i} & = S v_{i} +  \lambda I_{M_{\alpha}}v_{i} \\
                                      & = \mu_{i} v_{i} + \lambda v_{i}          \\
                                      & = (\mu_{i} + \lambda)v_{i}
\end{align*}
Thus, the eigenvalues of $S + \lambda I_{M_{\alpha}}$ are $\mu_{1} + \lambda, \ldots, \mu_{M_{\alpha}} + \lambda$, so
\begin{align*}
    |S + \lambda I_{M_{\alpha}}| = \prod_{i=1}^{M_{\alpha}} (\mu_{i} + \lambda)
\end{align*}
Therefore,
\begin{align*}
    \log{|A|} & = \log{|H_{\tilde\theta\tilde\theta}|} + \log{\prod_{i=1}^{M_{\alpha}} (\mu_{i} + \lambda)} \\
              & = \log{|H_{\tilde\theta\tilde\theta}|} + \sum_{i=1}^{M_{\alpha}} \log{(\mu_{i} + \lambda)}
\end{align*}
Differentiating the Laplace-approximated marginal likelihood in Equation \eqref{eq:laplace_log_marginal_likelihood_lambda} with respect to $\lambda$, with $\hat{\alpha}$ and $\mu_{i}$ evaluated at the current MAP estimate, yields the following empirical Bayes updating equation.
\begin{align*}
    \frac{\partial \log{p(\mathcal{D}|\lambda)}}{\partial \lambda} \approx \frac{M_{\alpha}}{2\lambda} - \frac{1}{2} \hat{\alpha}^\top \hat{\alpha} - \frac{1}{2} \sum_{i=1}^{M_{\alpha}}\frac{1}{\mu_{i} + \lambda} & = 0 \\
    M_{\alpha} - \lambda ||\hat{\alpha}||^2 - \left(M_{\alpha} - \sum_{i=1}^{M_{\alpha}}\frac{\mu_{i}}{\mu_{i} + \lambda}\right)                                                                                     & = 0 \\
    \sum_{i=1}^{M_{\alpha}}\frac{\mu_{i}}{\mu_{i} + \lambda} - \lambda ||\hat{\alpha}||^2                                                                                                                            & = 0
\end{align*}
Let
\begin{align*}
    g(\lambda) = \sum_{i=1}^{M_{\alpha}}\frac{\mu_{i}}{\mu_{i} + \lambda} - \lambda \|\hat{\alpha}\|^2
\end{align*}
For the precision matrix $A$ to be positive definite at the local maximum of the posterior distribution, it is necessary that $\mu_i + \lambda > 0$ for all $i$. Therefore, letting $\mu_{\min}$ be the minimum value of $\mu_i$, the domain of $\lambda$ is $\lambda > \max(0, -\mu_{\min})$. When $\mu_{\min} < 0$, as $\lambda \to (-\mu_{\min})^{+0}$ at the left end of the domain, $g(\lambda) \to -\infty$, and also as $\lambda \to \infty$, $g(\lambda) \to -\infty$. Furthermore, the derivative of $g(\lambda)$,
\begin{align*}
    g'(\lambda) = - \sum_{i=1}^{M_{\alpha}} \frac{\mu_{i}}{(\mu_{i} + \lambda)^2} - \|\hat{\alpha}\|^2
\end{align*}
can change sign multiple times due to the mixture of signs of $\mu_i$. As a result, there can be multiple (0, 2, 4, $\dots$) solutions (stationary points) satisfying $g(\lambda) = 0$ within the interval.
\begingroup
\singlespacing
\setlength{\bibsep}{0pt}
\bibliographystyle{apalike}
\bibliography{references}
\endgroup

\end{document}